\documentclass[sigplan, 10pt]{acmart}

\copyrightyear{2019} 
\acmYear{2019} 
\setcopyright{acmlicensed}
\acmConference[EuroSys '19]{Fourteenth EuroSys Conference 2019}{March 25--28, 2019}{Dresden, Germany}
\acmBooktitle{Fourteenth EuroSys Conference 2019 (EuroSys '19), March 25--28, 2019, Dresden, Germany}
\acmPrice{15.00}
\acmDOI{10.1145/3302424.3303957}
\acmISBN{978-1-4503-6281-8/19/03}

\usepackage{kotex}
\usepackage{graphicx}
\usepackage{balance}  % for  \balance command ON LAST PAGE  (only there!)
\usepackage{afterpage}
\usepackage{xspace}
\usepackage{tabularx}
\usepackage{multirow}
\usepackage{makecell}
\usepackage{array}
\usepackage{caption}
\usepackage[labelformat=simple]{subcaption}
\usepackage{makecell}
\usepackage{booktabs}
\usepackage{hyperref}
\hypersetup{pdftex,colorlinks=true,allcolors=black,pdfpagemode=UseOutlines}
\usepackage{hypcap}

\newcommand{\tool}{Parallax\xspace}
\usepackage[symbol]{footmisc}
\usepackage{color}

\newcommand{\thickhline}{%
	\noalign {\ifnum 0=`}\fi \hrule height 1pt
	\futurelet \reserved@a \@xhline
}
\newcolumntype{"}{@{\hskip\tabcolsep\vrule width 1pt\hskip\tabcolsep}}
\newcolumntype{P}[1]{>{\centering\arraybackslash}p{#1}}
\newcommand{\eat}[1]{}

\usepackage{listings}
\usepackage{multicol}

\DeclareFixedFont{\ttb}{T1}{txtt}{bx}{n}{8} % for bold
\DeclareFixedFont{\ttm}{T1}{txtt}{m}{n}{8}

\newcolumntype{x}[1]{>{\centering\arraybackslash\hspace{0pt}}p{#1}}
\definecolor{deepblue}{rgb}{0,0,0.5}
\definecolor{deepred}{rgb}{0.6,0,0}
\definecolor{deepgreen}{rgb}{0,0.5,0}
\definecolor{gray}{rgb}{0.33,0.33,0.33}
\newcommand\pythonstyle{\lstset{
		language=Python,
		basicstyle=\ttm,
		commentstyle=\color{deepgreen},
		keywords={def,return,self,with,as,for,in},
		keywordstyle=\ttb\color{deepblue},
		emphstyle=\ttb\color{deepred},    % Custom highlighting style
		frame=tb,                         % Any extra options here
		showstringspaces=false,            %
numbers=left,
xleftmargin=2.5em,
framexleftmargin=1.5em
}}

\lstnewenvironment{python}[1][]
{
\pythonstyle
\lstset{#1}
}
{}

\begin{document}

\lstset{language=Python}

\title[\tool: Sparsity-aware Data Parallel Training of DNNs]{\tool: Sparsity-aware Data Parallel Training of Deep Neural Networks}

\renewcommand*{\thefootnote}{\fnsymbol{footnote}}
\Urlmuskip=0mu plus 1mu

\author{Soojeong Kim}
\affiliation{
	\institution{Seoul National University}
}
\email{soojeong\_kim@snu.ac.kr}
\author{Gyeong-In Yu}
\affiliation{
	\institution{Seoul National University}
}
\email{gyeongin@snu.ac.kr}
\author{Hojin Park}
\affiliation{
	\institution{Seoul National University}
}
\email{hojinpark.cs@gmail.com}
\author{Sungwoo Cho}
\affiliation{
	\institution{Seoul National University}
}
\email{sungwoocho@snu.ac.kr}
\author{Eunji Jeong}
\affiliation{
	\institution{Seoul National University}
}
\email{ejjeong@snu.ac.kr}
\author{Hyeonmin Ha}
\affiliation{
	\institution{Seoul National University}
}
\email{hyeonmin.ha@snu.ac.kr}
\author{Sanha Lee}
\affiliation{
	\institution{Seoul National University}
}
\email{sanhaleehana@snu.ac.kr}
\author{Joo Seong Jeong}
\affiliation{
	\institution{Seoul National University}
}
\email{joosjeong@snu.ac.kr}
\author{Byung-Gon Chun}
\authornote{Corresponding author}
\affiliation{
	\institution{Seoul National University}
}
\email{bgchun@snu.ac.kr}

\renewcommand{\shortauthors}{S. Kim, et al.}

\begin{abstract}
The employment of high-performance servers and GPU accelerators for training deep neural network models have greatly accelerated recent advances in deep learning (DL).
DL frameworks, such as TensorFlow, MXNet, and Caffe2, have emerged to assist DL researchers to train their models in a distributed manner.
Although current DL frameworks scale well for image classification models, there remain opportunities for scalable distributed training on natural language processing (NLP) models.
We found that current frameworks show relatively low scalability on training NLP models due to the lack of consideration to the difference in sparsity of model parameters.
In this paper, we propose \tool, a framework that optimizes data parallel training by utilizing the sparsity of model parameters. 
\tool introduces a hybrid approach that combines Parameter Server and AllReduce architectures to optimize the amount of data transfer according to the sparsity.
Experiments show that \tool built atop TensorFlow achieves scalable training throughput on both dense and sparse models while requiring little effort from its users.
\tool achieves up to 2.8x, 6.02x speedup for NLP models than TensorFlow and Horovod with 48 GPUs, respectively.  
The training speed for the image classification models is equal to Horovod and 1.53x faster than TensorFlow.
\end{abstract}




\maketitle
\renewcommand*{\thefootnote}{\arabic{footnote}}
\setcounter{footnote}{0}

\section{Introduction}
\label{sec:intro}

It is a common practice nowadays for deep learning (DL) practitioners to utilize a cluster of GPU resources for training deep neural networks.
This is mainly motivated by the fact that recent deep neural network architectures involve very large computations~\cite{inception, resnet, nmt} and are trained on large datasets~\cite{imagenet, lm1b_dataset}, typically requiring multiple GPUs in order to finish training within a reasonable time limit.
There are a few parallelization strategies for accelerating training on multiple GPUs: running multiple model replicas that process disjoint datasets (data parallelism), partitioning a single model among multiple devices (model parallelism), and a mixture of the previous two strategies (hybrid parallelism).
Among these techniques, data parallelism is the most widely used thanks to its simplicity~\cite{fb1hour, inception, lm1b}, and is supported by most DL frameworks such as TensorFlow~\cite{tensorflow}, PyTorch~\cite{pytorch}, MXNet~\cite{mxnet}, Caffe2~\cite{caffe2}, and Horovod~\cite{horovod}, to increase training throughput by processing data in parallel.

There are a number of recent works that push the limit of data parallel training~\cite{fb1hour, ibm48min, prefernet15min, tencent6min}, achieving near-perfect throughput scaling efficiency\footnote{Scaling efficiency measures the percentage of speedup (in terms of throughput) in distributed training compared to the ideal, linear speedup when the same amount of GPUs are used.} of 99.2\% with thousands of GPUs~\cite{tencent6min}.
However, all of these works focus on parallelizing image classification models.
Little attention has been paid to training models from other domains, namely natural language processing (NLP) models.
In fact, we observed that using TensorFlow~\cite{tensorflow} to train NMT~\cite{nmt} and LM~\cite{lm1b} -- NLP models for neural machine translation and language modeling, respectively -- with 48 GPUs leads to scaling efficiencies of only 19.0\% and 7.0\% (Section~\ref{sec:evaluation}).
Current solutions to data parallel training are inadequate for handling a certain characteristic of these NLP models: sparsity of model parameters.

Multi-dimensional arrays that hold the parameters of a DL model can be classified into \textit{dense variables} and \textit{sparse variables}\footnote{We use the term \textit{variable}, following TensorFlow.
A \textit{sparse/dense variable} is different from a \textit{sparse/dense array}, which has its own mathematical meaning regarding the number of nonzero elements.}, depending on how their elements are accessed.
For a dense variable, all elements are accessed at least once during a single training iteration.
On the other hand, for a sparse variable, only a subset of the elements are accessed in one iteration.
Image classification models, such as the Inception-V3~\cite{inception} model, usually consist solely of dense variables for convolutional layers and fully connected layers.
We refer to such models as \textit{dense models}.
In contrast, NLP models have both dense variables and sparse variables.
For instance, the aforementioned LM~\cite{lm1b} model uses dense variables for internal long short-term memory (LSTM) cell parameters and sparse variables for word embeddings.
We define such models as \textit{sparse models}.

Sparse models tend to have larger variables than dense models, and must be dealt with differently in terms of parameter synchronization to maintain reasonable scalability.
For example, the largest variable in the dense model Inception-V3, weight of the fully connected layer, has 2.05 million elements, while the largest variable in the sparse model LM, the embedding matrix, has 406 million elements.
Synchronizing a large variable across multiple GPUs requires significant network bandwidth and consumes many CPU clocks for aggregating results from GPUs.
Thus, na\"ively communicating all elements of a large sparse variable, even though only a small subset is accessed, results in relatively low scalability.
At the same time, however, treating all variables as sparse variables is inefficient, as there are highly optimized implementations for communicating dense variables across GPUs such as the NCCL~\cite{nccl} library.

In this paper, we introduce \tool, a framework that takes the sparsity of variables into account to optimize data parallel training.
We analyze how the amount of data transfer changes according to whether variables are sparse or dense in two different training architectures: \textit{Parameter Server} and \textit{AllReduce}.
Based on this analysis, \tool pursues a hybrid approach that uses the Parameter Server architecture for handling sparse variables and the AllReduce architecture for handling dense variables.
Moreover, \tool partitions large sparse variables by a near-optimal number of partitions to maximize parallelism while maintaining low computation and communication overhead.
\tool further optimizes training with local aggregation and smart operation placement to mitigate communication overhead.
Graph transformation in \tool automatically applies all of these optimizations and the data parallel training itself at the framework level to minimize user efforts for writing and optimizing a distributed program by composing low-level primitives.

We have implemented \tool on top of TensorFlow~\cite{tensorflow} 1.6 with Horovod~\cite{horovod} 0.11.2.
Experiments on two sparse NLP models, LM~\cite{lm1b} and NMT~\cite{nmt}, and two dense image classification models, ResNet-50~\cite{resnet} and Inception-V3~\cite{inception}, show that \tool can speed up DL training of sparse models while achieving similar performance to state-of-the-art frameworks on dense models.
\tool achieves up to 2.8x and 6.02x speedup for the NLP models compared to TensorFlow and Horovod on 48 GPUs, respectively.
The training speed for the image classification models is equal to Horovod and 1.53x faster than TensorFlow.
Although we used NLP models for our evaluation to demonstrate the effectiveness of sparsity-aware data parallel training, \tool's techniques can be applied to any sparse model, such as speech recognition~\cite{speech, speechseq} and graph neural networks~\cite{graphnet}.
Even more, the performance gain is earned with absolutely no manual optimizations from the user -- merely a few lines of code are needed to use the \tool API.

The rest of the paper is organized as follows.
Section~\ref{sec:background} describes the DL background related to \tool and the motivation of utilizing model sparsity to optimize distributed training, while Section~\ref{sec:approach} introduces two sparsity-aware techniques of \tool.
Sections~\ref{sec:design} and \ref{sec:impl} present the design and implementation of \tool.
Section~\ref{sec:evaluation} presents evaluation results.
Section~\ref{sec:rltwk} presents related work and Section~\ref{sec:conc} concludes.

\begin{figure}[t]
\centering
  \begin{subfigure}[b]{0.46\columnwidth}
		\centering
		\includegraphics[width=0.9\linewidth]{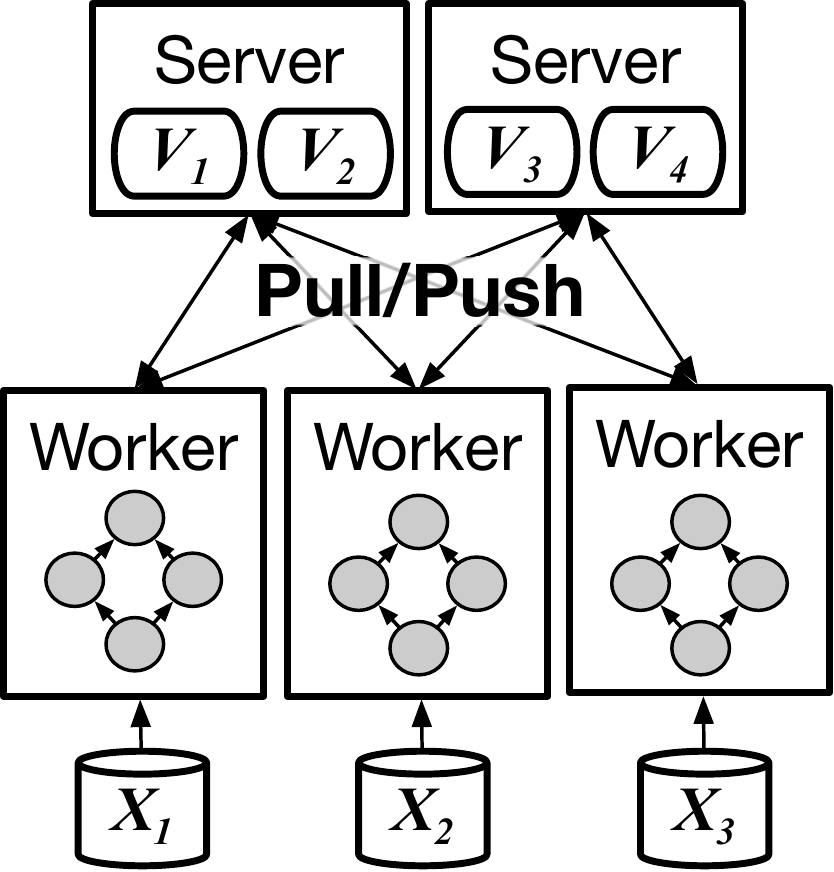}
		\caption{PS Architecture}
    \label{fig:general_ps}
	\end{subfigure}
  \begin{subfigure}[b]{0.53\columnwidth}
		\centering
		\includegraphics[width=0.9\linewidth]{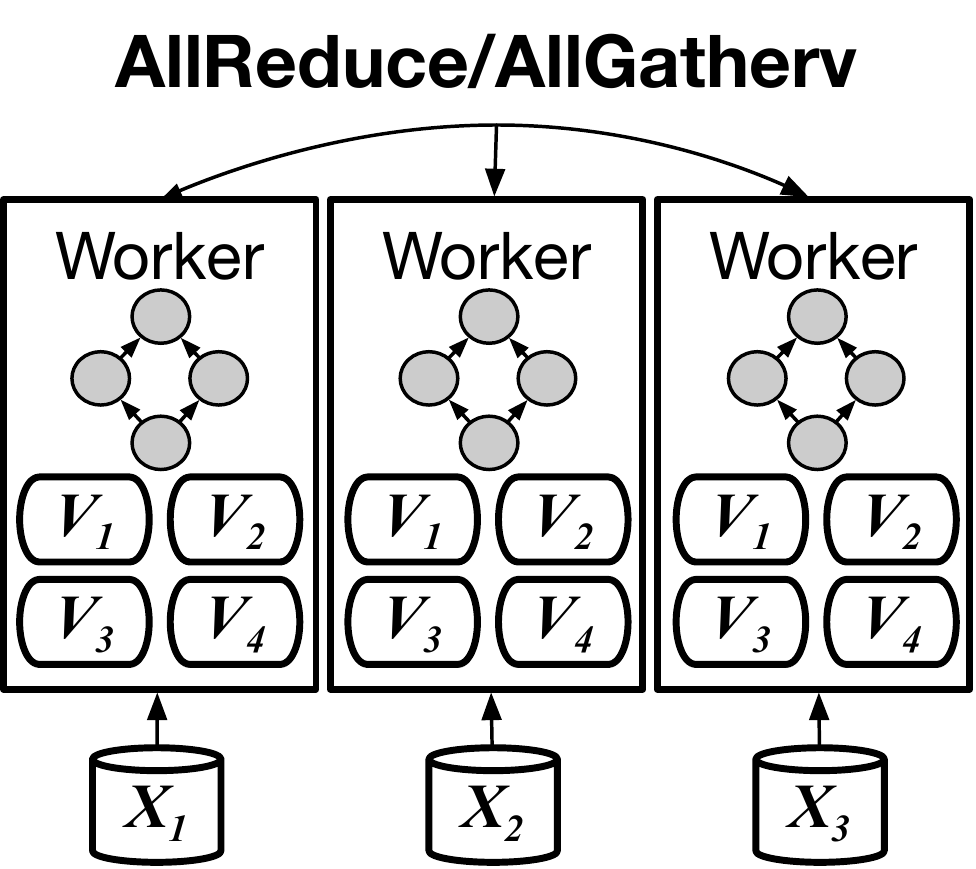}
		\caption{AllReduce Architecture}
    \label{fig:general_mpi}
	\end{subfigure}
  \caption{
  	The Parameter Server architecture and the AllReduce architecture.
  }
	\vspace{-4pt}
\end{figure}

\section{Background and Motivation}
\label{sec:background}
In this section, we briefly discuss data parallel distributed training and its two representative architectures: Parameter Server and AllReduce.
We also explain the motivation for taking model sparsity into account when training a DL model in a distributed manner.

\subsection{Data Parallel Distributed Training} \label{subsec:data_parallel}
A DL model refers to a neural network architecture, which is trained via gradient descent; the loss value of the model is calculated from forward computations, and the loss is passed back through the model according to the backpropagation algorithm to compute gradients.
These gradients are then used to update corresponding variables that compose the neural network.
Data parallel distributed training is utilized to process several mini-batches simultaneously with multiple GPUs.
GPUs are set to perform the same computation on different mini-batches, each producing a unique set of gradients.
In case of asynchronous training, the gradients from one GPU are used to update variables without waiting for other GPUs.
On the other hand, for synchronous training, all GPUs wait for one another to finish their gradient computation for variables.
Then, the computed gradients are aggregated before being used to update corresponding variables.
For both asynchronous and synchronous training, data communication between GPUs and machines is necessary to share the computed gradients.

For asynchronous training, the staleness of model variable updates is known to negatively impact the model's accuracy and produce relatively unpredictable results~\cite{revisitsync, staleness, tradeoff}.
Thus, many DL models are trained synchronously~\cite{fb1hour, nmt, vgg, pixelcnn}. This paper also assumes synchronous training, although we note that \tool supports both synchronous and asynchronous training.

\paragraph{Data Parallel Training Architectures}
Two widely-used data parallel distributed training architectures are the Parameter Server (PS)~\cite{ps_arch} architecture and the AllReduce (AR) architecture.
The PS architecture, initially proposed for topic modeling~\cite{ps_arch}, has been extensively used in previous works~\cite{tensorflow, mxnet, adam} thanks to the scalable structure that allows a large set of variables to be distributed into multiple machines.
A typical PS architecture consists of \textit{server} and \textit{worker} processes as described in Figure~\ref{fig:general_ps}.
Server processes store subsets of model variables (\textit{V\textsubscript{1}, ... , V\textsubscript{4}}) in memory, while worker processes \textit{pull} variables from servers to perform local computations on their respective mini-batches ($X\textsubscript{1}, X\textsubscript{2}, X\textsubscript{3}$) and later \textit{push} gradients with respect to variables back to servers.
As a result, variable synchronization between workers is done indirectly via server processes.

For the AR architecture, there is no process dedicated just for holding variables, as shown in Figure~\ref{fig:general_mpi}.
Rather, all workers are given a replica of variables and share locally computed gradients via collective communication primitives such as \texttt{AllReduce}~\cite{eff_allreduce, ib_allreduce} and \texttt{AllGatherv}~\cite{allgatherv}.
\texttt{AllReduce} reduces values from all processes to a single value, while \texttt{AllGatherv} simply gathers the values from all processes.
More formally, for the gradient $\frac{\partial L}{\partial v}(X_i)$ of a loss function $L$ with respect to a variable $v$ given a mini-batch data $X_i$, where worker $i$ processes $X_i$ ($i \in {1, ..., N}$), \texttt{AllReduce} aggregates gradients from all workers by computing the sum of gradients $\sum_{i=1}^{N}\frac{\partial L}{\partial v}(X_i)$.
On the other hand, \texttt{AllGatherv} aggregates gradients by concatenating the gradients into $[\frac{\partial L}{\partial v}(X_1), ..., \frac{\partial L}{\partial v}(X_N)]$.
Then, these primitives broadcast the aggregated gradients back to all processes.
The replica of variables housed in each worker is updated using the aggregated gradients, thereby all replicas in different workers are always synchronized.
This collective mechanism makes data parallel training simple because all workers always have the same variable values, thus there are no synchronization issues regarding variable updates.
Since the AR architecture is easier to use and shows better performance compared to the PS architecture for image classification models~\cite{horovod, perf_model}, recent attempts to scale out DL training~\cite{fb1hour, ibm48min, prefernet15min, tencent6min} employ AR as their distributed training architecture.

A major collective communication implementation used for the AR architecture is NCCL~\cite{nccl}, a well-known collective communication library that takes advantage of the GPU topology within and across multiple machines.
Depending on how GPUs are connected in a machine and across machines, NCCL composes different ring structures to achieve better performance.
It provides a highly optimized communication implementation, which is especially effective when the GPUs in the cluster support GPUDirect P2P or GPUDirect RDMA~\cite{gdr}.
Most DL frameworks that support distributed training, such as TensorFlow~\cite{tensorflow}, PyTorch~\cite{pytorch}, MXNet~\cite{mxnet}, Caffe2~\cite{caffe2} and Chainer~\cite{chainer}, adopt NCCL as their collective communication implementation.

\subsection{Necessity of Sparsity-awareness} \label{subsec:sparsity}
Although existing DL frameworks demonstrate scalable performance for data parallel training on large GPU clusters, their results are mostly based on well-known image classification models; there still remain untapped opportunities for scaling distributed training for models with sparse variables.
%A representative example of a dense variable would be a kernel matrix of a convolutional layer in convolutional neural networks (CNN).
%For each iteration, we utilize the entire matrix to produce outputs of the layer.
A representative example of a sparse variable would be an embedding matrix, which maps a word to an embedding vector.
Since sentences in a mini-batch typically include only a subset of an entire vocabulary list, only the corresponding rows of the embedding matrix is read and updated at each iteration.
For efficient memory management and computation, most DL frameworks provide special data structures for handling sparsity.
Instead of using a single array to represent sparse data such as a gradient of a sparse variable, two separate arrays are used -- one for the actual values, and another for indicating the value indices within the data, similar to the compressed sparse row (CSR) format~\cite{csr}.
For example, TensorFlow~\cite{tensorflow} manages dense data using a \texttt{Tensor} abstraction, while sparse data correspond to \texttt{IndexedSlices} or \texttt{SparseTensor} that contain two \texttt{Tensor}s to hold nonzero indices and values separately.

We claim that just like the data structures for sparse data, distributed data parallel training should also be aware of the different characteristics of dense and sparse variables.
To support this statement, we conducted experiments to show how the performance trend of training sparse models differs from that of dense models, regarding the underlying training architecture as well as partitioning variables for the appropriate architecture.
Moreover, this claim is further backed by researches from the machine learning community that employed data parallel training to train sparse models~\cite{lm1b, nmt}.

\begin{table}[t]
	\centering
	\begin{tabular}{lrrrrr}
		\toprule
		\multirow{2}[3]{*}{Models}
		& \multicolumn{2}{c}{\# Elements}  & \multirow{2}[3]{*}{$\alpha_{model}$}    & \multicolumn{2}{c}{Throughput}  \\
		\cmidrule(lr){2-3}                           \cmidrule(lr){5-6}
		& \multicolumn{1}{c}{Dense} & \multicolumn{1}{c}{Sparse} & & \multicolumn{1}{c}{PS} & \multicolumn{1}{c}{AR} \\
		\midrule
		ResNet-50      & 23.8M          & 0             & 1          & 5.8k                & \textbf{7.6k}                 \\
		Inception-v3   & 25.6M           & 0             & 1         & 3.8k                & \textbf{5.9k}                 \\
		LM          & 9.4M           & 813.3M          &  0.02        & \textbf{98.9k}     & 45.5k                         \\
		NMT            & 94.1M          & 74.9M           &  0.65         & \textbf{102k}      & 68.3k                         \\
		\bottomrule
	\end{tabular}
	\caption{
		The total size of dense and sparse variables, $\alpha_{model}$, and the training throughput (images or words per sec) of PS and AR architectures for four DL models, including two image classification models (ResNet-50, Inception-v3) and two NLP models (LM, NMT).
		The experiments are conducted on 48 GPUs using the cluster environment, datasets, and batch sizes described in Section~\ref{sec:evaluation}. 
		The PS column shows the results of TensorFlow using the PS architecture, and the AR column shows the results of Horovod using \texttt{AllReduce} for dense variables and \texttt{AllGatherv} for sparse variables.
	}
	\vspace{-10pt}
	\label{table:model_dataset}
\end{table}

\paragraph{Choosing Appropriate Training Architectures}
Table~\ref{table:model_dataset} shows that the sparsity of a model is an important factor when selecting a distributed training architecture.
It depicts the training throughput of four DL models along with their variable sizes and a ratio factor $\alpha_{model}$ that describes how sparse the model parameters are.
$\alpha_{model}$ is a weighted sum of $\alpha$ values of variables in the model, where the weight of each variable is proportional to its number of elements.
We define the $\alpha$ value of a variable as the average ratio of the number of elements that are actually used by a worker in one iteration to the total number of elements.
The first two models in the table, ResNet-50 and Inception-v3, are dense models and thus they do not contain sparse variables.
The next two models, LM and NMT, are sparse models, containing both dense and sparse variables.

Results show that the AR architecture is preferable for dense models, while the PS architecture performs better for sparse models.
This is because different distributed training architectures use network bandwidth in different ways; we discuss this further in Section~\ref{subsec:hybrid}.
To the best of our knowledge, no prior work considers the sparsity of models when selecting the distributed training architecture.

\paragraph{Impact of Partitioning Sparse Variables}
When using the PS architecture, it is common to partition large variables into multiple pieces to overcome memory constraints or to reduce load imbalance between server processes.
However, even when the memory requirements are satisfied and there is no significant load imbalance present, the number of partitions of sparse variables can affect overall performance.

Table~\ref{table:partitioning} shows the throughput of training the sparse models, LM and NMT, on various numbers of sparse variable partitions using the PS architecture.
Although all cases satisfy memory constraints and avoid significant load imbalance, the performance improvement for using the best possible choices (128 and 64 partitions for LM and NMT, respectively) and the worst possible choices (8 partitions for both models) is meaningful for both models; 1.98x for LM and 1.12x for NMT.
It is also worth noting that blindly increasing the number of partitions is not optimal, as the throughput at 256 partitions is worse than at 128 partitions in the LM model.
The performance improvement comes from the parallelization of operations for sparse variables; we describe the reasons for speedup in detail in Section~\ref{subsec:partitioning}.

\begin{table}[t]
	\centering
	\begin{tabular}{crrrrrr}
		\toprule		
		\multirow{2}[3]{*}{Model} & \multicolumn{6}{c}{\# Partitions} \\		
		\cmidrule(lr){2-7} 		
		& \multicolumn{1}{c}{8} & \multicolumn{1}{c}{16} & \multicolumn{1}{c}{32} & \multicolumn{1}{c}{64} & \multicolumn{1}{c}{128} & \multicolumn{1}{c}{256}\\
		\midrule		
		LM & 50.5k                 & 78.6k                  & 96.5k                  & 96.1k                  & 98.9k           & 93.2k\\	
		\midrule		
		NMT & 90.7k                 & 97.0k                 & 96.5k                 & 101.6k                 & 98.5k                 & 100.0k	\\
		\bottomrule   
	\end{tabular}
	\caption{Training throughput (words/sec) according to the number of partitions for LM and NMT models, using the PS architecture.
		The experiment setup is the same as Table~\ref{table:model_dataset}.}
	\label{table:partitioning}
	\vspace{-10pt}
\end{table}
\section{Sparsity-aware Data Parallel Training} \label{sec:approach}
Motivated by the experiment results in Section~\ref{subsec:sparsity}, we propose two sparsity-aware techniques to improve the performance of distributed training for sparse models: 1) a hybrid architecture of PS and AR, and 2) automatic searching of the optimal number of sparse variable partitions. % with low searching overhead.

\subsection{Hybrid Architecture} \label{subsec:hybrid}

\begin{figure*}[t]
	\centering
	\begin{subfigure}{0.3\textwidth}
		\centering
		\includegraphics[width=1.0\textwidth]{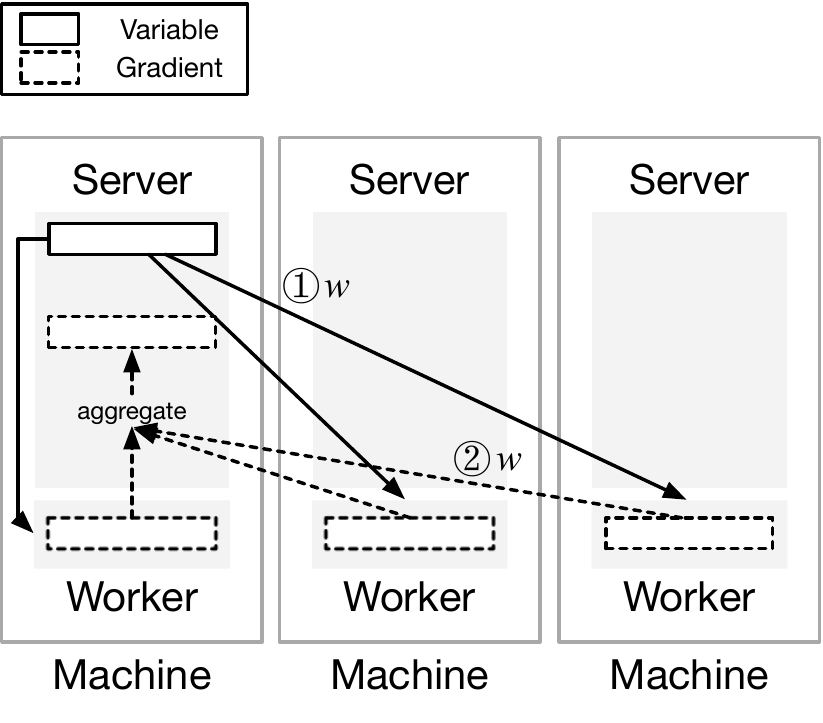}
		\caption{PS, dense variable and its gradient.}
		\label{fig:ps_dense_data_transfer}
	\end{subfigure} 
	\begin{subfigure}{0.3\textwidth}
		\centering
		\includegraphics[width=1.0\textwidth]{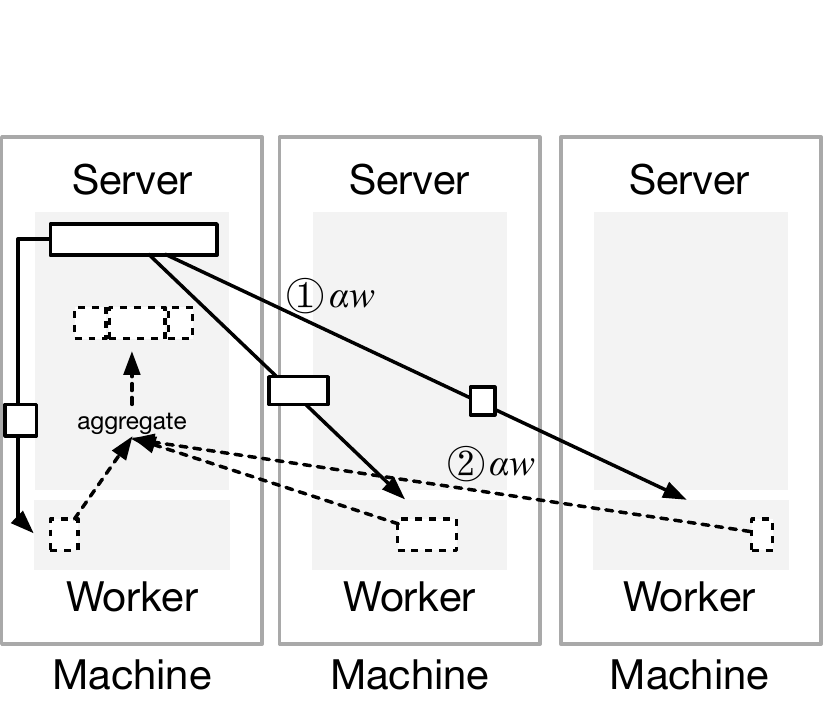}
		\caption{PS, sparse variable and its gradient.}
		\label{fig:ps_sparse_data_transfer}
	\end{subfigure}
	\begin{tabular}{c}
		\begin{subfigure}{0.33\textwidth}
			\centering
			\includegraphics[width=1.0\textwidth]{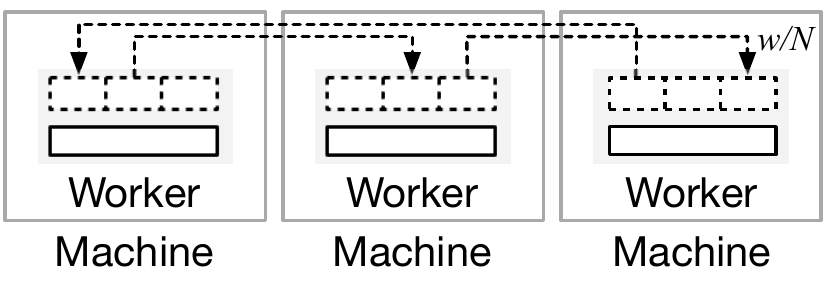}
			\caption{AR, gradient w.r.t. dense variable.}
			\label{fig:mpi_dense_data_transfer}
		\end{subfigure} \\
		\begin{subfigure}{0.33\textwidth}
			\centering
			\includegraphics[width=1.0\textwidth]{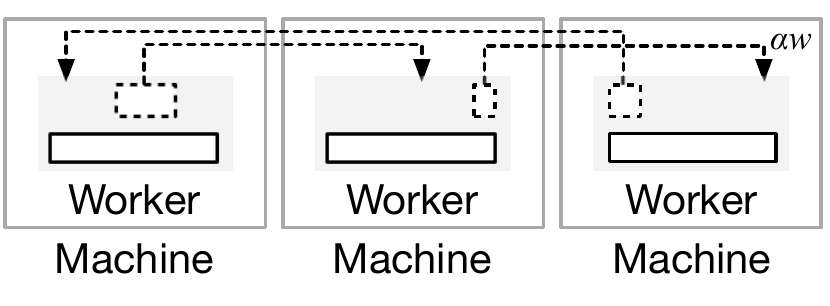}
			\caption{AR, gradient w.r.t. sparse variable.}
			\label{fig:mpi_sparse_data_transfer}
		\end{subfigure}
	\end{tabular}
	\vspace{-4pt}
	\caption{
		Data transfer for each type of variable and its gradient from/to a machine according to the training architecture.
		Bars with solid edge represent variables, while bars with dotted line represent gradients.
		Gradients of a dense variable from different workers are reduced by computing a sum of the gradients, while gradients of a sparse variable are aggregated by concatenating the arrays.}
	\vspace{-0pt}
	\label{fig:data_transfer}
\end{figure*}

As shown in Table~\ref{table:model_dataset}, the training speeds of the PS and AR architectures are affected by model sparsity.
The PS architecture performs faster when the model is sparse, while the AR architecture shows better performance when the model is dense.
We analyze this trend further by formulating the size of data transferred across the network during one training iteration for both architectures.

Figure~\ref{fig:data_transfer} shows how each training architecture synchronizes progress from multiple workers, for dense and sparse variables.
To simplify the explanation, we assume each machine contains only one worker process, and a server process is colocated with the worker in the PS architecture case.
Moreover, $w$ refers to the average size of variables as bytes, $N$ is the number of machines, and $\alpha$ is the element ratio of variables defined in Section~\ref{subsec:sparsity}.

Regarding a single dense variable in the PS architecture, a server process sends $w$ bytes to $N-1$ machines each, resulting in a network transfer of $w(N-1)$ bytes (Figure~\ref{fig:ps_dense_data_transfer}\textcircled{1}). 
The network cost for a server process occurs only for $N-1$ machines instead of all $N$ machines because server and worker processes in the same machine communicate locally within the machine without involving network communication.
Similarly, the server receives gradients of the same size back from the $N-1$ machines, leading to another $w(N-1)$ bytes (Figure~\ref{fig:ps_dense_data_transfer}\textcircled{2}).
Thus, for a single dense variable, the machine that houses the corresponding server sends and receives a total of $2w(N-1)$ bytes of data over the network for each iteration.
The network transfer for a sparse variable (Figure~\ref{fig:ps_sparse_data_transfer}) is similar to the dense variable case;
as defined in Section~\ref{subsec:sparsity}, each worker utilizes $\alpha$ of the elements in a sparse variable (in average), so they fetch $\alpha w$ bytes from the server and push back the same amount of gradients to the server.\footnote{We omitted the network transfer for exchanging nonzero indices since it is negligible in most cases compared to nonzero values.}
Therefore, the amount of network transfer for a sparse variable becomes $2 \alpha w(N - 1)$.

The data transfer behavior of the AR architecture varies depending on the actual algorithm implementation of \texttt{AllReduce} and \texttt{AllGatherv} being used.
Here, we assume the ring algorithm~\cite{ring}, one of the most popular collective communication algorithms, which is used in the NCCL library.
Regarding a single dense variable, each worker sends and receives $w/N$ bytes of data for $2(N-1)$ communication steps, where gradients are reduced for the first $N-1$ steps and the reduced values are broadcast back to all workers for the next $N-1$ steps.
Figure~\ref{fig:mpi_dense_data_transfer} shows the algorithm for one communication step, in which $2w/N$ bytes are going into and out of a single worker via network transfer.
Repeating this for $2(N-1)$ steps, we get a grand total of $4w(N-1)/N$ bytes for a machine.
On the other hand, for a sparse variable, each worker sends and receives $\alpha w$ bytes of data for $N-1$ communication steps in order to \texttt{AllGatherv} gradients for that variable (Figure~\ref{fig:mpi_sparse_data_transfer}), resulting in $2 \alpha w(N-1)$ bytes of network transfer for each machine.
The One Variable column of Table~\ref{table:data_transfer} summarizes these discussions about network transfer for a single variable, depicting all possible combinations of dense or sparse variables and the PS or AR architectures.

\begin{table}[t]
	\centering    
	\begin{tabular}{cccc}
		\toprule
		Type & Arch & One Variable & $m$ Variables \\
		\midrule
		\multirow{2}[3]{*}{Dense} & PS & $2w(N - 1)$ & $4wm \frac{N - 1}{N}$ \\
		\cmidrule(lr){2-4} & AR & $4w \frac{N - 1}{N}$ & $4wm \frac{N - 1}{N}$ \\
		\midrule
		\multirow{2}[3]{*}{Sparse} & PS & $2 \alpha w(N - 1)$ & $4 \alpha wm \frac{N-1}{N}$ \\
		\cmidrule(lr){2-4} & AR & $2 \alpha w(N - 1)$ & $2 \alpha wm(N - 1)$ \\
		\bottomrule
	\end{tabular}
	\caption{The amount of network transfer required per machine for each type of variable according to the training architecture.}
	\vspace{-10pt}
	\label{table:data_transfer}
\end{table}

Moving from one variable to multiple variables, we add additional assumptions about the variable distribution across servers.
We assume that all variables occupy the same amount of memory ($w$ bytes) and are distributed evenly across server processes.
In such a balanced PS architecture, each machine manages $\frac{m}{N}$ dense variables where $m$ is the number of dense variables in a model.
For the $\frac{m}{N}$ variables that a machine manages, a total of $2w(N-1) \times \frac{m}{N}$ bytes of network transfer occurs; for the other $m - \frac{m}{N}$ variables that the machine does not manage, $2w \times (m - \frac{m}{N})$ bytes of transfer occur since the machine needs to fetch $w$ bytes for each variable and send another $w$ bytes for each corresponding gradient.
Thus, the amount of network transfer per machine for $m$ dense variables becomes $2w(N-1) \times \frac{m}{N} + 2w \times (m - \frac{m}{N}) = 4wm \frac{N-1}{N}$.
Similarly, each machine requires $4 \alpha wm \frac{N-1}{N}$ bytes of network transfer in order to synchronize $m$ sparse variables.

Unlike the PS architecture, all variables in the AR architecture are housed by all workers, and thus are present in all machines.
Thus, we can simply derive the total amount of network transfer per machine by multiplying $m$ with the amount of network transfer for a single dense or sparse variable, giving us $4wm \frac{N - 1}{N}$ and $2 \alpha wm(N - 1)$ bytes, respectively.

Both PS and AR architectures require the same amount of network transfer for a machine, with $m$ dense variables.
However, the amount required for a single dense variable that is managed by the machine is much larger in the PS architecture. 
The machine that manages the variable needs to handle $2w(N - 1)$ bytes of network transfer, compared to $2w$ bytes of other machines.
This difference can possibly lead to a communication bottleneck in the machine in charge of the variable, while network bandwidth for other machines is under-utilized.
We anticipate this asymmetry between machines to be the root cause of the performance difference between PS and AR architectures.
Since a DL model comprises multiple layers and there are dependencies between them, pull and push requests for variables in different layers are scattered along the timeline.
On the other hand, there is no such asymmetric network transfer for the AR architecture, and therefore no particular machine becomes a bottleneck.
Recent studies~\cite{horovod, perf_model} show that the NCCL-based AR architecture achieves higher performance on dense models such as ResNet-50~\cite{resnet}, Inception-v3~\cite{inception}, and VGG-16~\cite{vgg}.

For sparse variables, exchanging gradients using the AR architecture requires much more data transfer compared to the PS architecture.
As $N$ becomes larger, the difference between the two architectures becomes more significant.

Based on the analysis, we propose a hybridization of the two architectures to achieve the best of both worlds.
\tool employs a hybrid architecture in which the AR architecture handles dense variables and the PS architecture handles sparse variables.
Each worker has a replica of dense variables, while separate server processes manage only sparse variables.
Note that if the $\alpha$ value of a sparse variable is close to 1, then it may be helpful to handle the variable as a dense variable and use \texttt{AllReduce}, even though it requires $\frac{1}{\alpha}$ times larger network transfer compared to the PS architecture.
In this case, $\alpha$ should be large enough to make the gain from efficient network utilization of the AR architecture surpass the overhead of extra network transfer.

\subsection{Partitioning of Sparse Variables} \label{subsec:partitioning}
As stated in Section~\ref{subsec:sparsity} and Table~\ref{table:partitioning}, partitioning sparse variables can affect training throughput.
The fact that the performance goes up as the number of partitions increases up to 128, without any significant load imbalance, implies that there is inevitably another factor that contributes to the improvement.

We found that partitioning sparse variables effectively parallelizes the aggregation of the corresponding gradients, as well as the variable update operations.
Gradient aggregation and update operations for sparse variables require iterating through nonzero indices one by one to accumulate values with the same index.
Partitioning a sparse variable parallelizes these operations by dividing incoming values and indices into disjoint sets, and thus enables the parallel execution of such operations.
Meanwhile, increasing the number of partitions introduces additional overhead for stitching the partial results from each partition into one tensor to be used as input for other operations~\cite{tensorflow}.
It is also accompanied with the overhead of managing each partition of the variable as separate arrays.
These aspects are related to not only the DL model itself, but also the hardware specification of the cluster and batch size; simple rule-based heuristics are not able to find a reasonable optimum for various conditions.

To capture these effects, we suggest a cost-based model that predicts iteration time as a function of the number of partitions $P$:
\begin{equation}
\label{eq:model}
iter\_time = \theta_0 + \theta_1 * \frac{1}{P} + \theta_2 * P
\end{equation}
Parameter $\theta_0$ represents the constant cost for fixed computation and communication, which does not change over $P$.
$\theta_1$ captures the cost that can be parallelized and amortized by increasing $P$, while $\theta_2$ represents the overhead incurred by partitioning sparse variables.

\tool collects data points required to fit Equation~\ref{eq:model} by performing actual training with different values for $P$, for a few iterations.\footnote{\tool runs 100 iterations and discards values from the first 50 iterations to eliminate startup cost.}
Then, we fit the equation using mean-squared error of the sampled iteration time and prediction.
In order to reduce the number of samples while maintaining high accuracy, \tool exploits the fact that Equation~\ref{eq:model} is a convex function of $P$.
Setting $P$'s initial sample point to be the number of machines, \tool collects the iteration time for $P$ while doubling the value until the iteration time starts to increase.
Next, \tool repeats the process while halving $P$, again until the iteration time starts to go up.
The critical point of the convex function is located between the minimum and maximum $P$s of the collected data, hence the cost model can predict the optimal $P$ without performing any extrapolation.
\begin{figure}[!tb]
\centering
\begin{python}
<@\textcolor{deepred}{import parallax}@>
		
# create a graph as distributed version
with single_gpu_graph:
  ds = input_files_dataset()
  <@\textcolor{deepred}{ds = parallax.shard(ds)}@>
  en_texts, de_texts = ds.get_input_data()
  
  <@\textcolor{deepred}{with parallax.partitioner():}@>
    emb_enc = get_variable(shape=[...])
    emb_dec = get_variable(shape=[...])
  loss = build_NMT_model(en_texts, de_texts, 
                         emb_enc, emb_dec)
  grads_and_vars = compute_grads(loss)
		
  opt = GradientDescentOptimizer(LR=0.1)
  train_op = opt.update(grads_and_vars)
		
<@\textcolor{deepred}{graph\_runner = parallax.get\_runner(}@>
  <@\textcolor{deepred}{single\_gpu\_graph,}@>
  <@\textcolor{deepred}{resource\_info\_file,}@>
  <@\textcolor{deepred}{parallax\_config)}@>

for i in range(num_iters):
  graph_runner.run(train_op)
\end{python}
\caption{Example code for training the NMT model in a distributed multi-GPU environment with \tool.
Red lines represent the necessary modifications for adding \tool:
\texttt{shard} for splitting the input data for data parallelism, 
\texttt{partitioner} for partitioning sparse variables,
and \texttt{get\_runner} for performing automatic parallelization.}
\vspace{-13pt}
\label{fig:parallaxexample}
\end{figure}

\section{System Design}\label{sec:design}
\tool is a sparsity-aware data parallelization framework built on TensorFlow~\cite{tensorflow}, a state-of-the-art DL framework.
\tool enables users to utilize distributed multi-GPU environments when they have a single-GPU computation graph (i.e., a deep learning model developed for training on a single GPU).
It guarantees transparency while keeping scalable performance using a hybrid architecture with optimally partitioned sparse variables.
For the transparency, users do not need to write new code for data parallel training that requires prior knowledge for training architectures and sparsity of variables.
Instead, the framework provides an API that receives a single-GPU computation graph as input and automatically transforms the graph into a multi-GPU, multi-machine computation graph.

\subsection{Programming Interface}\label{subsec:api}
\begin{sloppypar}
	\tool provides simple programming interfaces: \texttt{shard}, \texttt{partitioner}, and \texttt{get\_runner}.
	Unlike single-GPU training, input data must be divided into disjoint subsets to be processed by different GPUs for data parallel distributed training. 
	\tool helps this process with the \texttt{shard} API, which receives input data and splits the data into multiple subsets so that each GPU can read a unique subset.
	When exploration for optimal partitioning is required through \texttt{partitioner}, the variables within \texttt{partitioner} context are partitioned using an optimal number of partitions searched by \tool.
	\texttt{get\_runner} is the main interface that accepts a single-GPU graph as well as resource information including the IP addresses (or hostnames) of machines and GPU IDs, and an optional \tool configuration (\texttt{ParallaxConfig}) object specifying extra arguments if needed. The configuration includes whether to use local aggregation or not, a file path to save trained variables and aggregation methods for each type of variable indicating whether to compute the average of gradients for dense (or sparse) variables over all GPUs or to compute the sum instead. 
	
	We illustrate how to use the \tool API with a code snippet example for training the NMT~\cite{nmt} model, a DL model for language translation.
	Figure~\ref{fig:parallaxexample} shows code for training the NMT model on a GPU cluster.
	\tool requires three modifications compared to a corresponding single-GPU training code: splitting input data across GPUs (line 6), creating partitioned variables using \texttt{partitioner} (line 9), and creating \tool's \texttt{graph\_runner} instead of the original framework's.
	First, a graph object is declared, \texttt{single\_gpu\_graph}, which is followed by the logic for preprocessing input data, the loss function, the gradients from backpropagation, and the gradient descent method for updating the variables (lines 4-17).
	The input data must be split across GPUs for data parallelism, and this can be accomplished with the \texttt{shard} interface.
	The \texttt{ds} object in line 5 represents the whole input data, while the \texttt{ds} object returned by \texttt{shard} in line 6 is a unique subset of dataset for a model replica.
	Next, users can create partitioned variables using \texttt{partitioner} in line 9.
	\tool finds and applies the optimal partitioning for the variables (\texttt{emb\_enc} and \texttt{emb\_dec}).
	Note that each \texttt{partitioner} partitions variables into the same number of partitions.
	When the user wants to partition variables in different granularities, multiple \texttt{partitioner}s must be created and applied independently.
	Then, the computation graph is transformed to be executable on multiple GPUs through the \texttt{get\_runner} interface.
	In lines 19-22 and line 25, the \texttt{graph\_runner} object returned by the \texttt{get\_runner} interface should be used in place of the graph runner of the original framework, since it is not aware of the fact that the computation graph has been converted for a distributed multi-GPU environment.
\end{sloppypar}

In the existing frameworks~\cite {tensorflow, mxnet}, users must use different APIs for constructing computation graphs depending on whether the training is done on a distributed environment or only on a single GPU.
For example, a user that wants to train a model using TensorFlow’s PS architecture must be aware of two types of processes - server and worker - and insert mechanisms for gradient aggregation and synchronization.
Meanwhile, \tool lets users recycle almost the same single-GPU code for constructing computation graphs on distributed environments, allowing easier utilization of multiple GPUs.
We discuss this point further in Section~\ref{sec:rltwk}.

\subsection{Execution Model} \label{subsec:architecture}
We outline the overall execution model of \tool as follows.
After a client initiates a job with a computation graph and resource information, \tool analyzes the computation graph to construct hybrid architecture.
If the graph only contains dense variables, \tool launches workers as many as the number of GPUs.
On the other hand, if sparse variables are included in the graph,  \tool launches a server process for each machine and a worker process for each GPU. 
When the processes are launched, the number of partitions for sampling is passed to the workers.
Worker processes transform the input graph to a distribute version and run for a small number of iterations on the given resources. 
During the graph transformation step, \tool separates dense and sparse variables and creates a distributed graph for AR and PS architectures (if necessary). 
Then, each worker sends its execution time to the master process which collects execution time according to the number of partitions. 
This process is repeated until sampling for variable partitioning ends.
Finally, \tool executes the transformed graph with optimally partitioned sparse variables.
Next, we explain the details of graph transformation.

\subsection{Automatic Graph Transformation} \label{subsec:transformation}
\tool carries out the transformation process adhering to several specific rules systemically as a substitution of user's manual modifications from a single-GPU graph to a distributed version.
\tool builds transformation rules for AR and PS architectures while maintaining transparency, correctness and scalability, and these rules are combined for hybrid architecture.
Note that the transformation rules do not automate hyperparameter tuning to find optimal hyperparameters such as learning rate or batch size.
\tool uses hyperparameters that are given from the input graph.

\begin{figure}[t]
	\centering
	\includegraphics[width=0.95\linewidth]{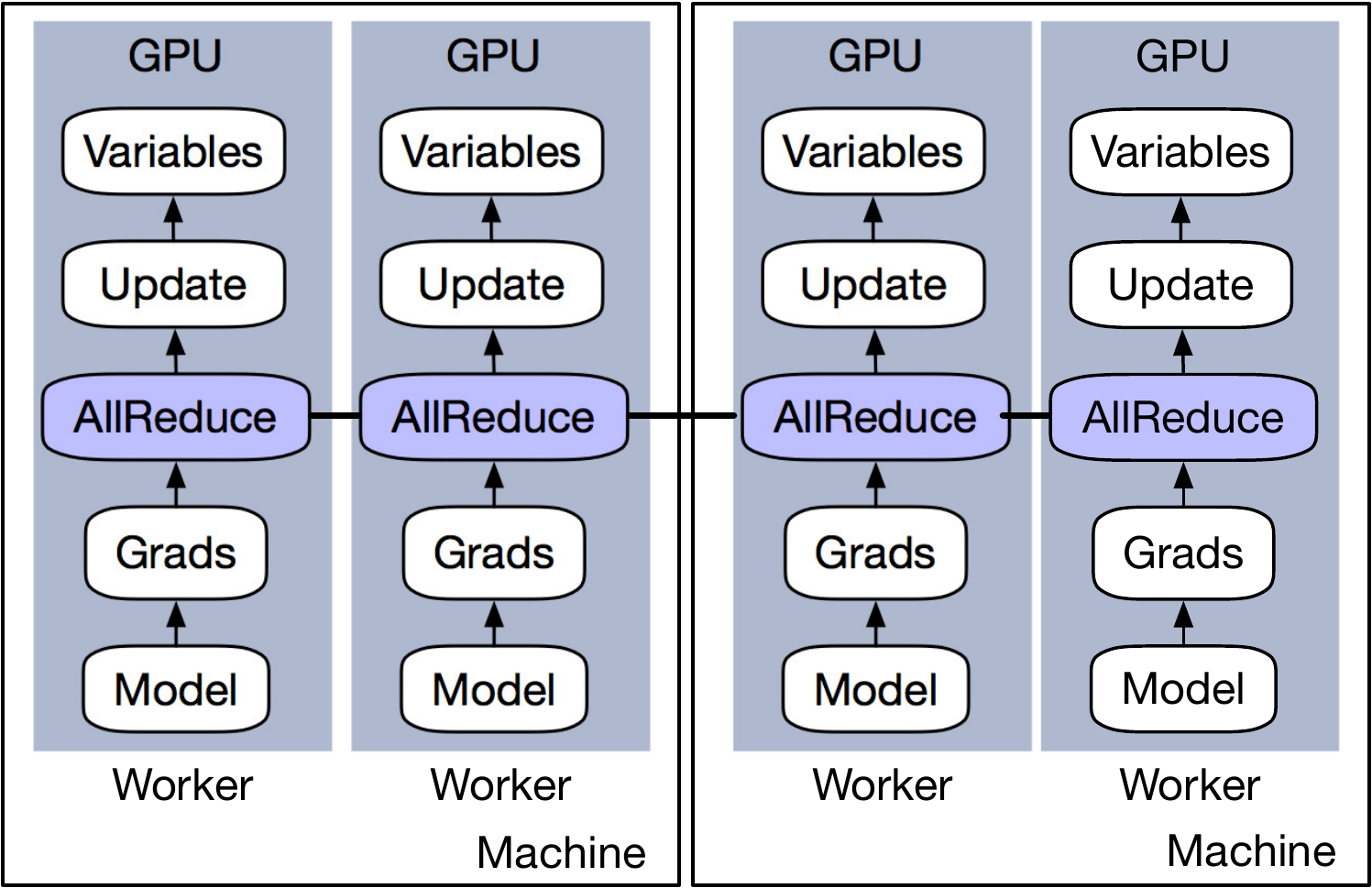}
	\caption{Graph transformation for AR architecture. In this example, the transformed graph uses \texttt{AllReduce} to aggregate gradients.}
	\label{fig:mpi}
\end{figure}

\paragraph{Transformation for AR} \label{subsubsec:mpi}
Figure~\ref{fig:mpi} shows graph transformation for AR architecture. It is relatively straightforward compared to the transformation for PS because each device carries individual copies of global states  (i.e., variables) and does not access states on the other devices.
\tool replicates all operations in the original single GPU graph and places a replica for each GPU in the resource specification.
The transformation is simple because of the homogeneity of all the processes (workers) that participate in training, unlike the PS architecture.
\tool automatically identifies gradients using information in a single-GPU graph to satisfy a transparent graph transformation.
To aggregate gradients across devices, \texttt{AllReduce} operations take place between operations that produce gradients using backpropagation (\texttt{Grads}) and their successors (\texttt{Models}).

\begin{figure}[t]
	\centering
	\includegraphics[width=\linewidth]{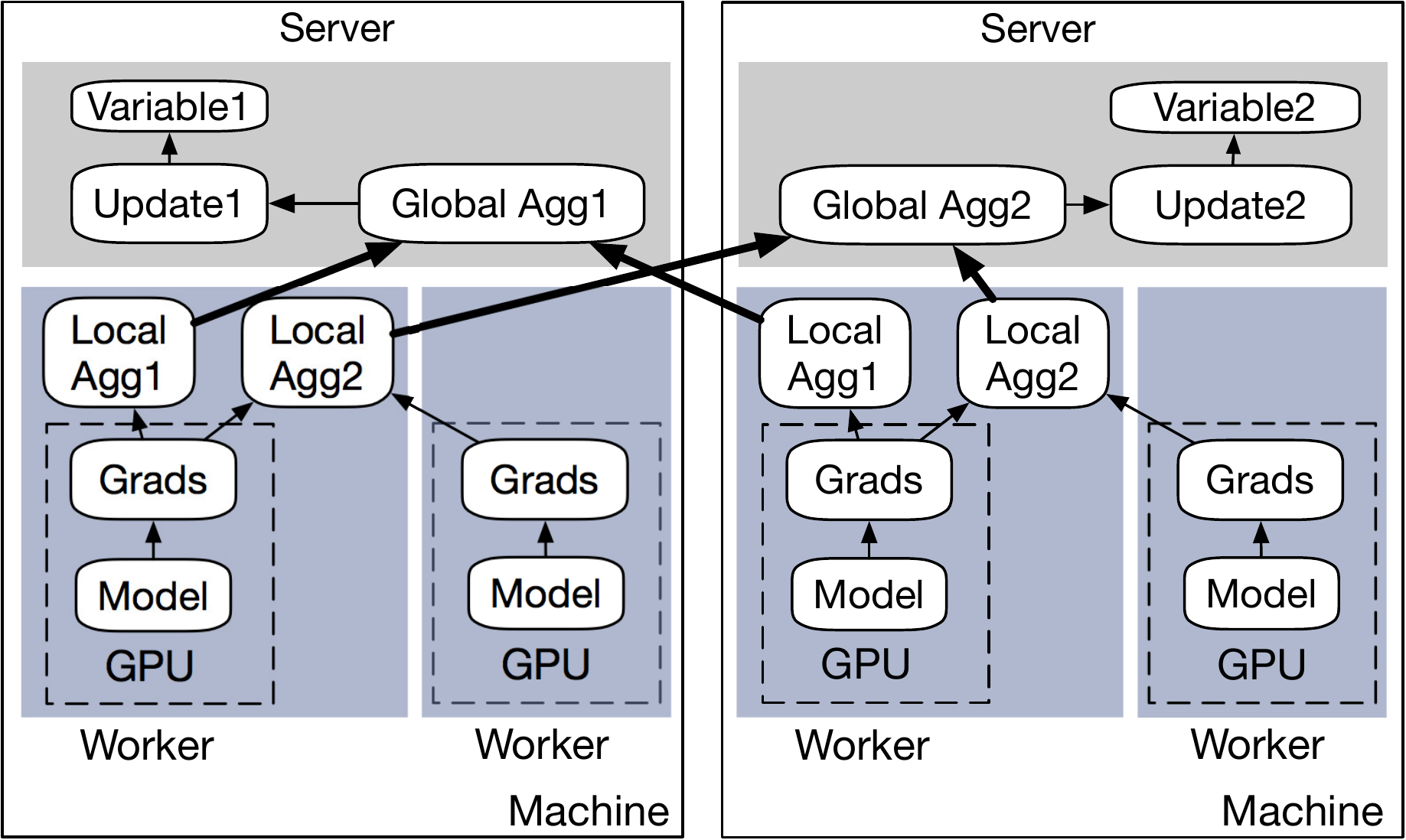}
	\caption{Graph transformation for PS architecture.}
	\vspace{-2.96pt}
	\label{fig:ps}
\end{figure}

\paragraph{Transformation for PS} \label{subsubsec:ps}
%\sloppypar{
\tool supports an optimized PS architecture using local aggregation and assigning operations effectively across machines.
Consequently, the graph transformation rules for PS are defined based on the optimized PS.
\tool transforms a single-GPU graph for PS architecture by creating a copy of forward and backward graph operations for each worker and distributing variables and their update operations across servers. \tool applies different replication and operation placement policies to variables, variable update operations, and main computation operations.
Figure~\ref{fig:ps} shows an example of the graph transformation. \tool launches a (parameter) server on each machine and a worker on each GPU in the given resource specification. This colocation of workers and a server in a machine works well since workers are GPU-intensive while servers run lightweight computation, which runs only on CPUs.
\tool evenly distributes variables (\texttt{VariableN}) across servers, and a large variable is partitioned to multiple pieces if the variable is specified as a partitioning target in the code.
	Each partitioned piece has a partitioned gradients aggregation and a partitioned update operation.
	\tool assigns update operations (\texttt{UpdateN}) in the same server with their variables to be updated.
	Identifying model variables and their update operations is feasible because DL frameworks~\cite{tensorflow, theano, mxnet, caffe2} treat them differently from mathematical operations, such as add or multiplication.
	Main computation operations that are used to compute gradients are replicated as many as the number of GPUs. \texttt{Model} and \texttt{Grads} represent operations for forward computation and backpropagation, respectively.
	Along with the detection of gradients, \tool identifies main computation operations by searching all the ancestor operations from the gradients in the graph.
	Gradients from each GPU are aggregated twice using aggregation operations for GPUs within a machine (\texttt{LocalAggN}) and between machines (\texttt{GlobalAggN}).
	The local aggregation reduces the amount of data communication between workers and servers, which is more expensive than communication between GPUs in the same machine. 
	The outputs of \texttt{GlobalAggN} are used to update model variables. \tool places a global aggregation operation (e.g., \texttt{GlobalAgg1}) on the same server with the variable (e.g., \texttt{Variable1}) to minimize data transfer between machines.

\begin{figure}[t]
	\centering
	\includegraphics[width=\linewidth]{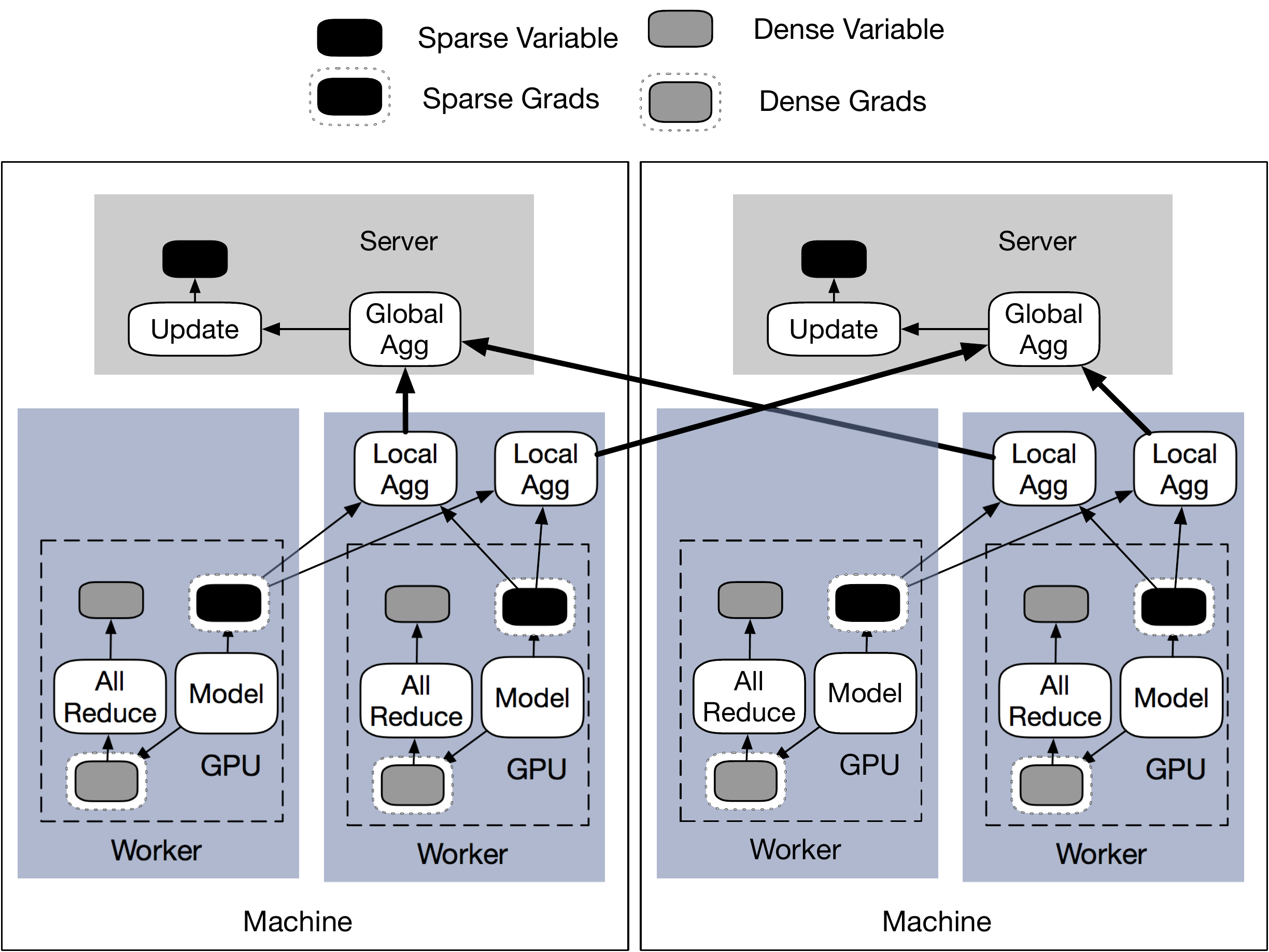}
	\caption{Graph transformation for hybrid architecture.}
	\vspace{-4pt}
	\label{fig:hybrid}
\end{figure}

\paragraph{Transformation for Hybrid} \label{subsubsec:hybrid}
Figure~\ref{fig:hybrid} shows the transformed graph for hybrid architecture.
Regardless of architectures, main computations (\texttt{Models} and \texttt{Grads}) are replicated in each GPU.
Then, \tool separates dense and sparse variables using the different data structures to handle gradients of each type.
Finally, a sparse variable follows PS transformation rules while AR transformation rules are applied to a dense variable.
The sparse variables are shared via server processes and global aggregation methods are inserted between locally aggregated gradients from each machine and update operations.
The dense variables replicated in each worker are updated using the aggregated gradients from \texttt{AllReduce}.
Because each variable is synchronized independently, applying different rules to each type of variables completes graph transformation for hybrid architecture. 

\section{Implementation}
\label{sec:impl}
We implemented \tool on TensorFlow~\cite{tensorflow} v1.6  with ~\texttt{AllReduce} operation using NCCL in Horovod~\cite{horovod} v0.11.2.
We implemented the graph transformation and distributed execution in Python. 

\paragraph{Identifying the sparsity of a variable} 
In TensorFlow, dense and sparse variable are distinguished by the different types of their gradient tensors.
The type is determined when the gradient tensor is generated by automatic differentiation, depending on how the variable is used in the forward computation.
For example, TensorFlow creates a sparse type gradient tensor for a variable used in a sparse access operation, \texttt{gather}.
\tool uses this type information to identify if a variable is either sparse or dense.

\paragraph{Graph transformation}
Graph transformation of \tool consists of inserting gradient aggregation operations for sparse variables and placing operations to specific resources.
Placing operations can be done with the \texttt{tf.device} API.
However, aggregating gradients requires additional steps as follows.
We first place accumulators on servers to aggregate the gradients of sparse variables, where each accumulator handles gradients of a single sparse variable.
When gradients are aggregated in an accumulator, a worker asks the server to read the aggregated gradient from the accumulator and update the corresponding variable.

To provide correct variable updates as done in a single-GPU code, \tool ensures that only one worker, namely a chief worker, triggers the operations for reading aggregated gradients and updating variables.
The other workers wait until these variable update operations are finished.
The chief's notification arrives through shared queues on each worker.
If the other workers also need aggregated gradients to trace their status during training or to compute a global norm of gradients for clipping, \tool changes the worker-side graphs to read the aggregated gradients from the variables where the chief worker saves them temporarily after reading from accumulators.
In case of local aggregation, \tool adds additional accumulators to each machine, and a worker in the machine becomes a local chief worker to collect gradients within a machine and send them to servers.

In addition, we modified the TensorFlow core to store gradients information, which is the result of auto-differentiation for model variables, in ~\texttt{MetaGraphDef} protobuf in TensorFlow.
The modified \texttt{MetaGraphDef} enables \tool to track exact mapping between model variables and their gradients.
\tool uses this information for inserting gradient aggregation operations.
\begin{figure*}[ht]
	\centering
	\begin{subfigure}[b]{0.33\textwidth}
		\centering
		\includegraphics[width=1.0\linewidth]{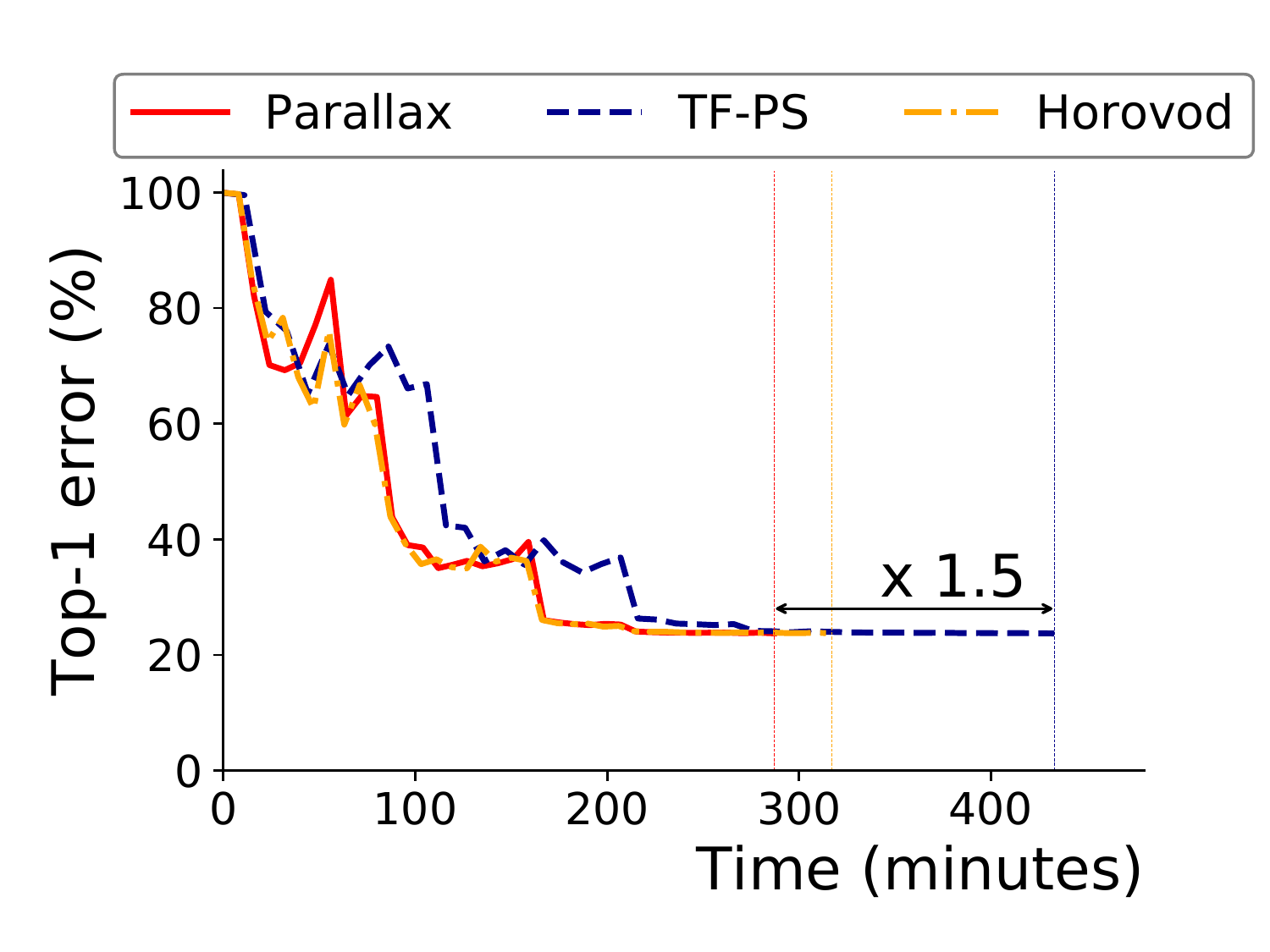}
		\caption{ResNet-50}
		\label{fig:resnet50_convergence}
	\end{subfigure} 
	\begin{subfigure}[b]{0.33\textwidth}
		\centering
		\includegraphics[width=1.0\linewidth]{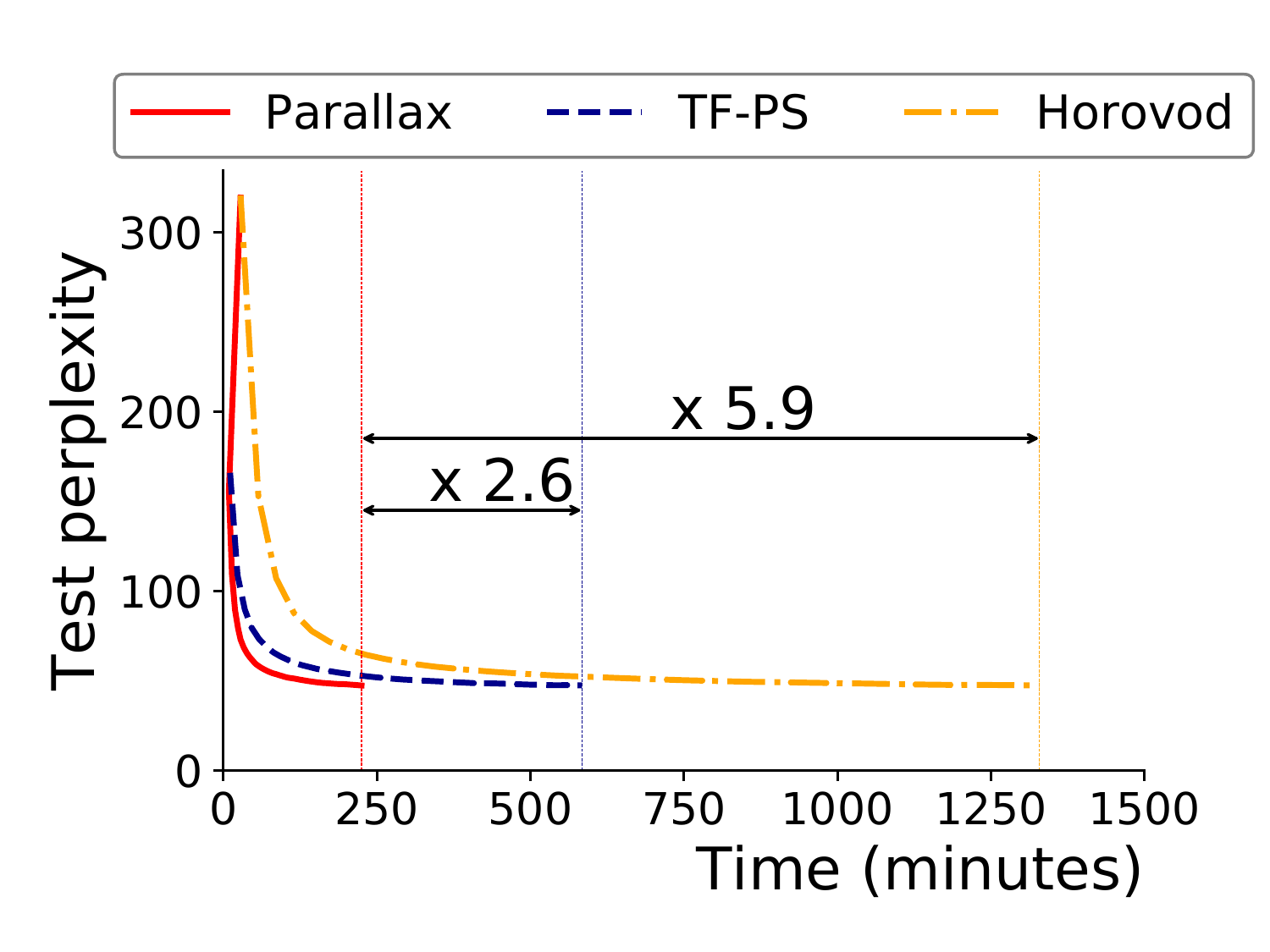}
		\caption{LM}
		\label{fig:lm1b_convergence}
	\end{subfigure}
    \begin{subfigure}[b]{0.33\textwidth}
	  \centering
	  \includegraphics[width=1.0\linewidth]{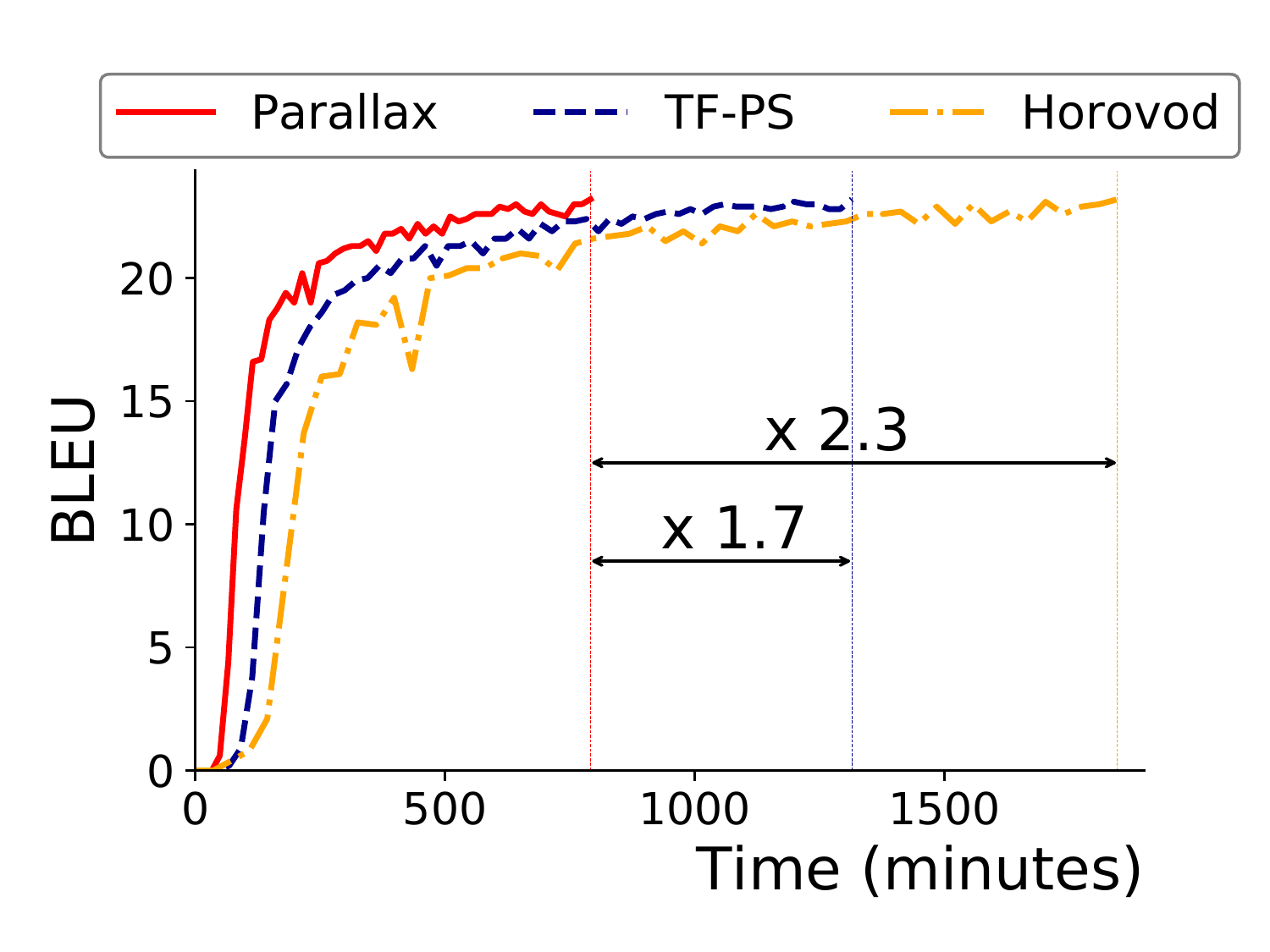}
	  \caption{NMT}
	  \label{fig:nmt_convergence}
    \end{subfigure}
	\caption{Convergence results. 
		(a) Top-1 validation error of ResNet-50.
		(b) Test perplexity of LM.
	    (c) BLEU score of NMT. \\
        The vertical lines represent where each framework reaches the same target value for Top-1 validation error, test perplexity, or BLEU score.
        The target values are 23.74\% for ResNet-50, 47.5 for LM, and 22.5 for NMT.}
	\label{fig:convergence}
\end{figure*}

\section{Evaluation}\label{sec:evaluation}
We evaluate \tool with experiments to answer the following questions:
\begin{itemize}
	\item Does \tool correctly transform computation graphs and improve convergence speed using the sparsity-aware data parallel training? (Section~\ref{subsec:convergence})
	\item Does \tool scale out well to multiple GPUs and machines? (Section~\ref{subsec:throughput})
	\item How much performance benefits do \tool's optimization techniques provide? (Sections~\ref{subsec:throughput}, ~\ref{subsec:hybrid_eval} and \ref{subsec:partitioning_eval})
	\item How does \tool's performance change under various sparsity degrees? (Section~\ref{subsec:sparsity_eval})
\end{itemize}

\subsection{Experiment Setup}\label{subsec:exp_setup}
\textbf{Cluster Configuration.}
We conducted all the experiments on a GPU cluster of 8 machines.
Each machine is equipped with two 18-core Intel Xeon E5-2695 @ 2.10 GHz processors with 256 GB RAM and 6 NVIDIA GeForce TITAN Xp GPU cards.
The machines are connected via Mellanox ConnectX-4 cards with 100Gbps InfiniBand. They run Ubuntu 16.04, CUDA 9.0, cuDNN 7, OpenMPI v3.0.0, and NCCL v2.1.

\textbf{Frameworks.}\label{subsec:eval_frame}
As baselines, we selected TensorFlow v1.6 as a representative DL framework for the PS architecture, and Horovod~\cite{horovod} v0.11.2 on TensorFlow for the AR architecture.
In the evaluation, TF-PS denotes TensorFlow with PS.
We let Horovod use NCCL for \texttt{AllReduce} since NCCL provides highly-optimized communication between GPUs compared to OpenMPI.
However, we inevitably use OpenMPI for \texttt{AllGatherv}, which is not provided by NCCL.

\textbf{Models and Datasets.}
We trained two image classification models and two NLP models in our experiments.
ResNet-50~\cite{resnet} and Inception-v3~\cite{inception}, are trained with the ImageNet (ILSVRC 2012)~\cite{imagenet} dataset that has 1.28M training images and 50K validation images in 1000 categories. 
LM~\cite{lm1b} is a language model that learns a probability distribution over sequences of words in a language.
It consists of a single layer of LSTM with hidden state of size 2048, projected to a 512-dimensional embedding.
We trained the LM model on the One Billion Word Benchmark~\cite{lm1b_dataset} that contains one billion words with the vocabulary size of 800K.
NMT~\cite{nmt} is a machine translation model, composed of 8-layer LSTMs of 1024 units with a bidirectional encoder of 1024-dimensional embedding.
We used the WMT English-German dataset~\cite{wmt} that has 4.5M sentence pairs for NMT model training.
As described in Table~\ref{table:model_dataset}, the image models are dense models, which consist of only dense variables, while the NLP models are sparse models, which contain both dense and sparse variables.
The batch size per GPU is 64 for ResNet-50 and Inception-v3, and it is 128 for LM and NMT.

\begin{figure*}[!htb]
	\centering
	\begin{subfigure}[b]{0.49\textwidth}
		\centering
		\includegraphics[width=0.9\textwidth]{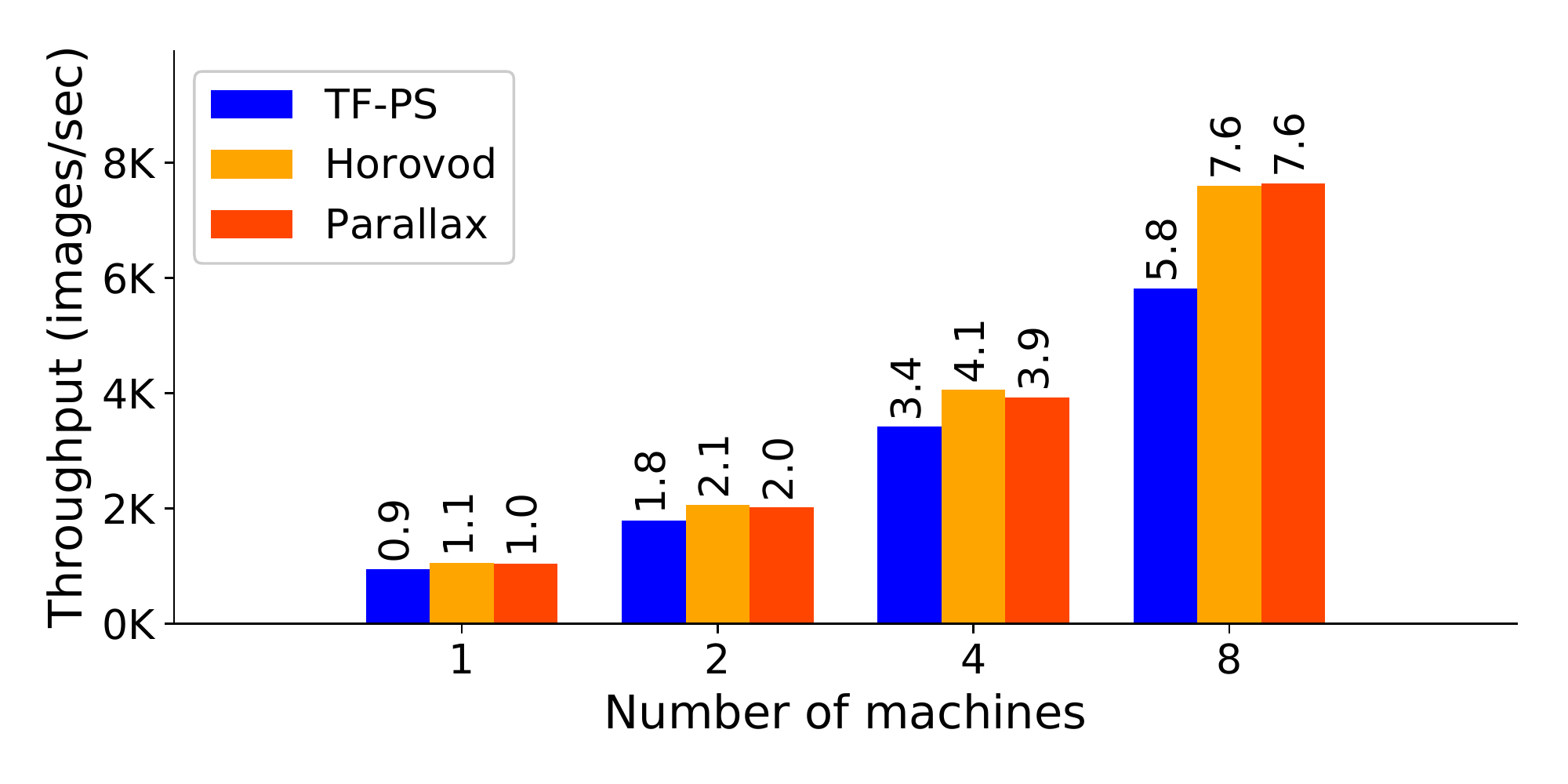}
		\caption{ResNet-50}
		\label{fig:resnet50_thp_cmp_manual}
	\end{subfigure} 
	\begin{subfigure}[b]{0.49\textwidth}
		\centering
		\includegraphics[width=0.9\textwidth]{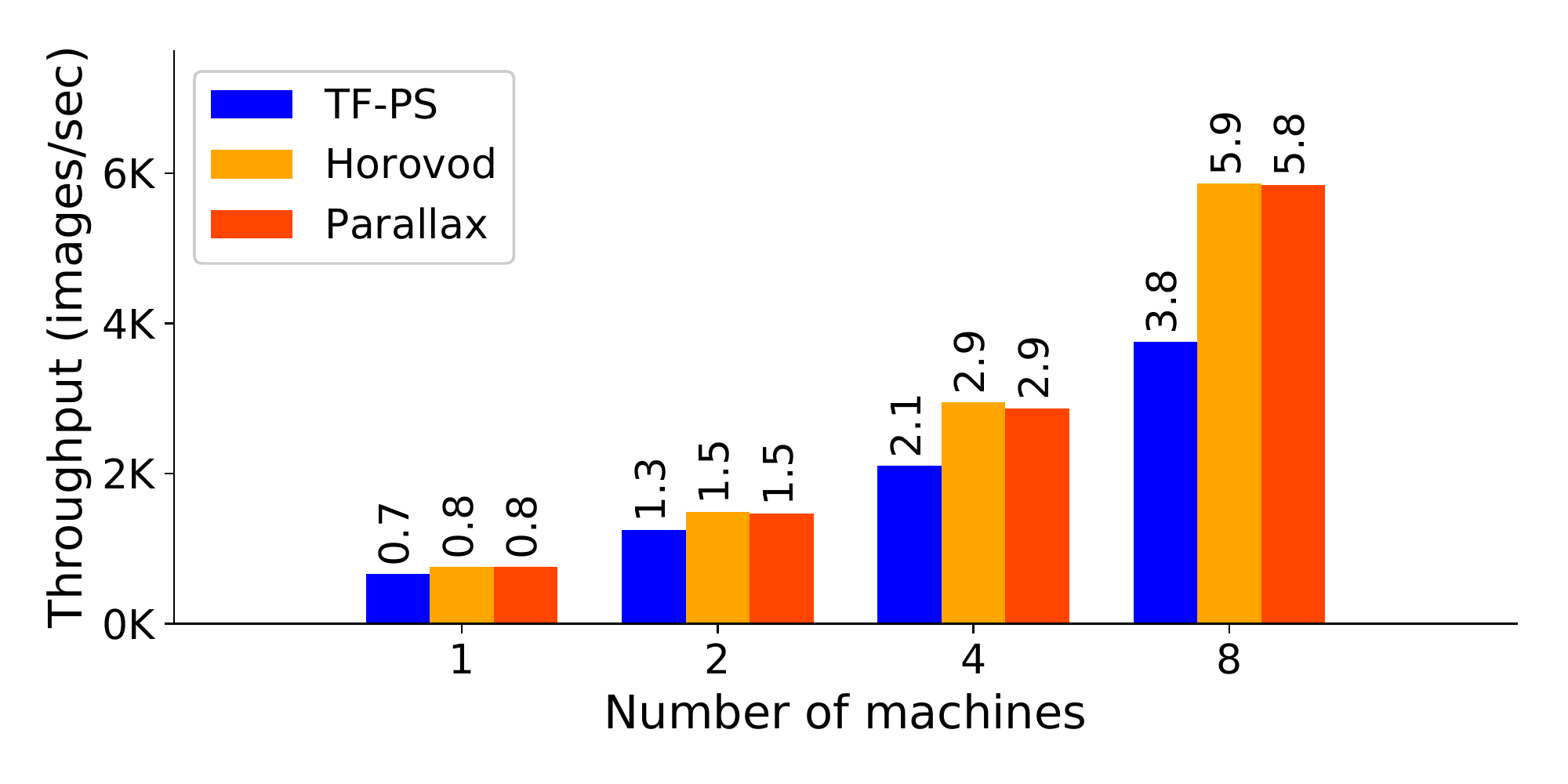}
		\caption{Inception-v3}
		\label{fig:inception3_thp_cmp_manual}
	\end{subfigure}
	\begin{subfigure}[b]{0.49\textwidth}
		\centering
		\includegraphics[width=0.9\textwidth]{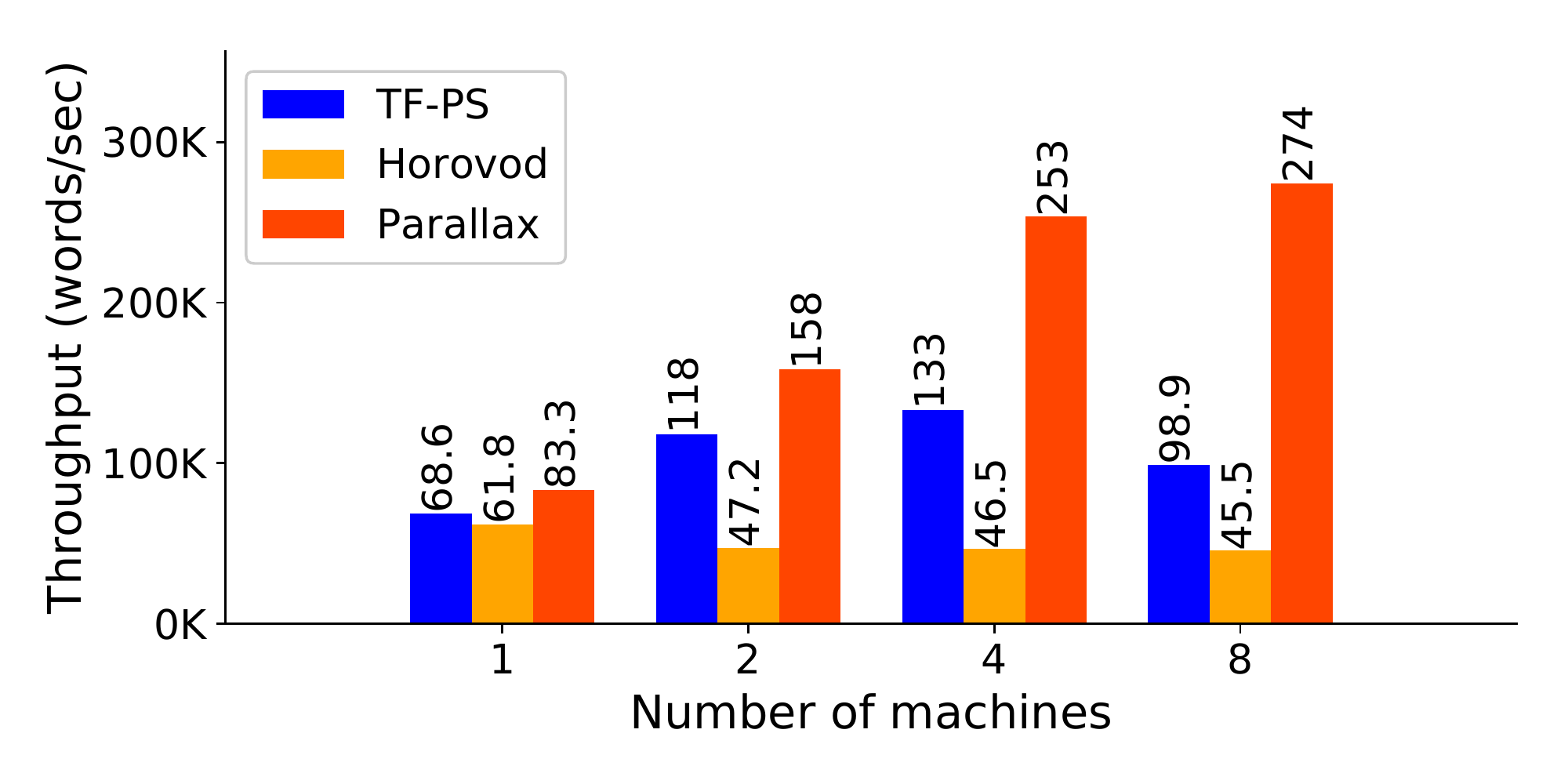}
		\caption{LM}
		\label{fig:lm1b_thp_cmp_manual}
	\end{subfigure} 
	\begin{subfigure}[b]{0.49\textwidth}
		\centering
		\includegraphics[width=0.9\textwidth]{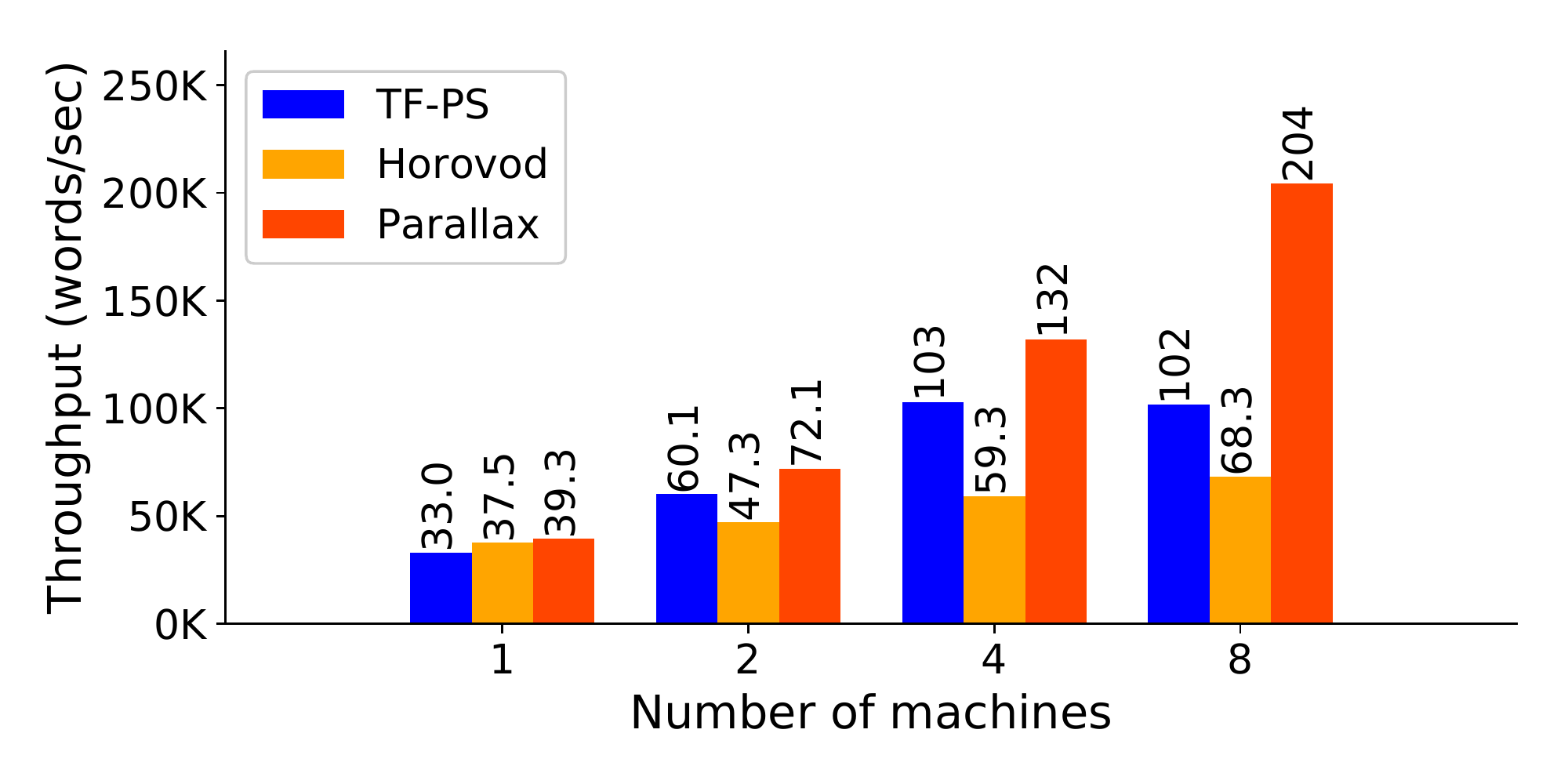}
		\caption{NMT}
		\label{fig:nmt_thp_cmp_manual}
	\end{subfigure}
	\caption{Training throughputs of (a) ResNet-50, (b) Inception-v3, (c) LM and (d) NMT on \tool, TF-PS and Horovod, varying the number of machines from 1 to 8.
		In dense models (ResNet-50 and Inception-v3), \tool outperforms TF-PS and shows the same performance with Horovod.
		In sparse models (LM and NMT), \tool is faster than both TF-PS and Horovod.}
	%\vspace{-4pt}
	\label{fig:thp_cmp_manual}
\end{figure*}

\subsection{Model Convergence}\label{subsec:convergence}
\tool correctly converges models as other frameworks, and the convergence speed is faster than or equal to TF-PS and Horovod.
Figure~\ref{fig:convergence} shows the convergence graphs of ResNet-50, LM, and NMT models.
We compare the training time taken for each framework to converge models, which is indicated by a model-specific metric reaching the same target values.
The target values are 23.74\% top-1 error for ResNet-50 experiments (Figure~\ref{fig:resnet50_convergence}), perplexity of 47.5 for LM experiments (Figure~\ref{fig:lm1b_convergence}), and BLEU score of 23.2 for NMT experiments (Figure~\ref{fig:nmt_convergence}).
ResNet-50, LM, and NMT experiments use 48, 36, and 24 GPUs, respectively.

The convergence speed in Figure~\ref{fig:convergence} demonstrates the relationship between the training architecture and the sparsity of models.
For example, ResNet-50 results confirm our findings that the AR architecture (Horovod) is efficient for the training of dense models than the PS architecture (TF-PS).
Horovod’s training takes less time than TF-PS for the same top-1 validation error.
\tool shows almost equal performance with Horovod because \tool utilizes only the AR architecture for dense models by using Horovod \texttt{AllReduce} operations.
The slight difference in convergence times of \tool and Horovod is due to random variable initialization and data shuffling effects unrelated to the techniques described in this paper.

On the other hand, TF-PS is faster than Horovod for the LM model as we expected.
For all LM model experiments, \tool automatically finds a near-optimal number of partitions for sparse variables using its regression-based method.
In the case of TF-PS and Horovod, we perform a manual search for the number of partitions as the frameworks do not provide automatic search mechanisms.
Thanks to Parallax’s hybrid architecture and optimizations such as local aggregation, Parallax achieves a 2.6x speedup compared to TF-PS and a 5.9x speedup compared to Horovod.

Similar to the LM experiments, the NMT model experiments were conducted after applying partitioning of sparse variables for each framework.
\tool converges 2.3x faster than Horovod and 1.7x faster than TF-PS.

\subsection{Performance and Scalability}\label{subsec:throughput}
Next, we show the performance of \tool by comparing the training throughput of \tool against those of TF-PS and Horovod.
Then, we evaluate the scalability of \tool as we increase the number of GPUs.

\paragraph{Training Throughput}
Figure~\ref{fig:thp_cmp_manual} shows the training throughput of \tool, TF-PS and Horovod.
According to Figures~\ref{fig:resnet50_thp_cmp_manual} and~\ref{fig:inception3_thp_cmp_manual}, Horovod achieves higher throughput compared to TF-PS on the dense models.
For these models, \tool achieves throughput similar to Horovod.
In contrast to the dense models, the three frameworks have significant performance differences for the sparse models.
Figures~\ref{fig:lm1b_thp_cmp_manual} and~\ref{fig:nmt_thp_cmp_manual} depict training throughput for LM and NMT.
On 48 GPUs, \tool shows 2.8x speedup and 2.0x speedup for LM and NMT compared to TF-PS, respectively.
Throughout all combinations of the number of machines and different DL models, \tool always outperforms or gives performance equal to both TF-PS and Horovod.

\begin{figure}[!h]
	\centering
	\includegraphics[width=0.99\linewidth]{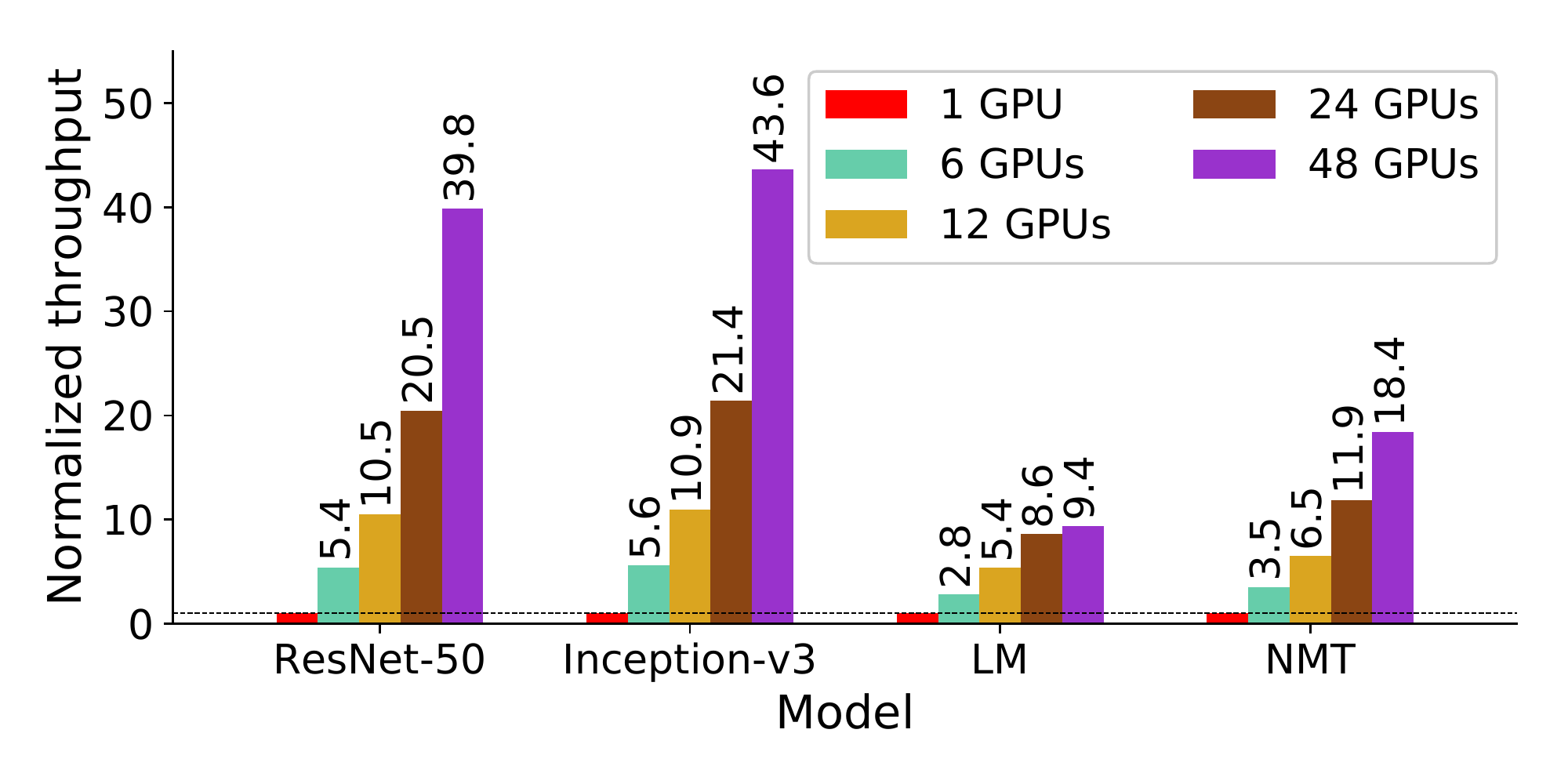}
	\caption{Scalability of \tool for each model. With 48 GPUs, normalized throughput of \tool is 39.8, 43.6, 9.4 and 18.4 for ResNet-50, Inception-v3, LM and NMT respectively, which scales better than TF-PS (30.4, 28.6, 3.4, 9.1) and Horovod (39.8, 44.4, 1.6, 6.1)}
	\vspace{-12pt}
	\label{fig:parallax_scalability}
\end{figure} 

\textbf{Scalability of \tool.}
Figure~\ref{fig:parallax_scalability} presents the scalability of \tool for the four models.
We define normalized throughput for $n$ GPUs ($n$ = 6, 12, 24, 48) as the ratio of the throughput for $n$ GPUs over the throughput for 1 GPU.
Ideally, the normalized throughput should be equal to the number of GPUs. 
The difference between ideal throughput and actual throughput comes from the added communication overhead for distributed training.
For ResNet-50 and Inception-v3, \tool scales out well, achieving 39.8x and 43.6x speedups on 48 GPUs. 
The scalability of LM and NMT is worse than ResNet-50 and Inception-v3. The normalized throughput of LM is 9.4, and that of NMT is 18.4 with 48 GPUs. 
LM and NMT stress more communication than ResNet-50 and Inception-v3 due to the large size of variables and relatively light computation.
For example, the number of variable elements exchanged per GPU is 101 million for NMT, 26 million for ResNet-50, and 24 million for Inception-v3. 

\begin{table}[t]
	\centering
	\begin{tabular}{lcccc}
		\toprule
		\multirow{1}{*}{\makecell{Models}} & \multirow{1}{*}{\makecell{AR}}   & \multirow{1}{*}{\makecell{Na\"ivePS}}     &
		\multirow{1}{*}{\makecell{OptPS}}    & \multirow{1}{*}{\makecell{HYB (AR + OptPS)}} \\
		\midrule
		LM      &  45.5k    &    98.9k    &   250k   &     274k \\
		NMT     &   68.3k    &   102k     &   116k   &  204k  \\
		\bottomrule
	\end{tabular}
	\caption{Training throughput (words/sec) of various architectures.}
	\label{table:hybrid_effect}
	\vspace{-14pt}
\end{table}

\subsection{Effect of Hybrid Architecture}\label{subsec:hybrid_eval}
To analyze the effectiveness of the hybrid architecture compared to employing only one architecture, we compare the throughputs of AR using \tool, the na\"ive PS architecture (Na\"ivePS) using TF-PS, optimized PS (OptPS) in \tool, and the hybrid architecture (HYB) based on AR and OptPs, as shown in Table~\ref{table:hybrid_effect}.
OptPS includes local aggregation and smart operation placement across server and worker processes.
We experiment for LM and NMT models using 8 machines with 48 GPUs.
In the experiments, sparse variable partitioning is applied to all architectures because of the large size of the sparse variables.

As we show in the previous section, Na\"ivePS (TF-PS) outperforms AR on sparse models - the speedup is 2.2x for LM and 1.5x for NMT.
OptPS improves the throughput of Na\"ivePS by 2.5x and 1.1x for LM and NMT, respectively. 
The speedup continues on HYB - it is 1.1x faster than OptPS for the LM model and 1.8x faster for the NMT model. 
HYB's performance improvement is more significant in the NMT model which has a similar ratio of sparse and dense variables (56\% of total variables are dense and the remaining 44\% are sparse).
On the other hand, the speedup of the LM model is relatively low as we progress from OptPS to HYB.
The reason is that the majority of variables in the LM model are sparse variables (the size of sparse variables is 99\% of the size of total variables), and the effect of optimizing the communication of dense variables by combining AR and PS is rather small.

\subsection{Sparse Variable Partitioning}\label{subsec:partitioning_eval}
	We present the efficiency of the sparse variable partitioning method of \tool for LM and NMT in Table~\ref{table:partition_eval}.
	The efficiency is measured by comparing throughput of \tool's method with that of a brute-force method that finds the optimal number of partitions by first starting from the smallest number of partitions possible without memory exceptions (4 and 2 partitions for LM and NMT, respectively) and gradually increasing the number of partitions by 2 to get better throughput.
	The brute-force method stops searching when the number of partitions is too large that throughput drops more than 10\% compared to the highest throughput observed.
	Compared to the results using the smallest number of partitions without exceeding the memory bound (Min), \tool's partitioning method improves the throughput by 2.84x and 1.64x for LM and NMT, respectively.
	Moreover, \tool's method does not fall behind more than 5\% compared to the brute-force method (Optimal).
	The brute-force method is much more inefficient than \tool; \tool spends at most 20 minutes to get sampling results of at most 5 runs while the brute-force method needs to collect results from more than 50 runs.

\begin{table}[t]
	\centering
	\begin{tabular}{lccc}
		\toprule
		Models & \tool  & Min & Optimal \\
		%\cmidrule(lr){4-5}
		%&&& Max & Min \\
		\midrule
		LM &   274k   &   96.5k   &  260.3k \\  
		\midrule
		NMT &  204k & 124.1k & 208k \\
		\bottomrule
	\end{tabular}
	\caption{
			Training throughputs (words/sec) from different partitioning methods with 8 machines (48 GPUs).
			The \tool column corresponds to \tool's partitioning method, the Min column shows the results of using the smallest number of partitions possible without memory exceptions, and the Optimal shows the results of the brute-force method.
	}
	\label{table:partition_eval}
	%\vspace{-2pt}
\end{table}

\begin{table}[t]
	\centering
	\begin{tabular}{lcccc}
		\toprule
		length & $\alpha_{model}$ & \tool & TF-PS & Speedup \\
		\midrule
		120	& 1.0& 437k &	214k &	2.04x \\
		60	&	0.52 &	511k & 219k & 2.33x \\
		30	&	0.28 &	536k &	221k & 2.43x\\
		15	& 0.16 &	557k &	193k &	2.89x \\
		8	&	0.1 &	480k &	159k &	3.02x \\
		4 & 0.07 &	285k &	94k &	3.03x \\
		1 & 0.04 &  82k & 24k & 3.42x \\
		\bottomrule
	\end{tabular}
	\caption{
			The training throughput (words/sec) of \tool and TF-PS, and speedup of \tool compared to TF-PS under various sparsity degrees ($\alpha_{model}$). 
		    \textit{length} represents the number of words in a data instance.}
	\label{table:sparsity}
	\vspace{-10pt}
\end{table}

\subsection{Effect of Sparsity Degree}\label{subsec:sparsity_eval}
	Table~\ref{table:sparsity} compares the training throughput (words/sec) under various sparsity degrees ($\alpha_{model}$) using Parallax and TF-PS.
	All experiments were performed on 48 GPUs using a constructed LM model that uses dense variables and vocabulary smaller than those of the original LM model to test under a wide range of $\alpha_{model}$ values.
	$\alpha_{model}$ is controlled by the number of words (length) in a data instance with the batch size fixed.
	The longer the length of a data instance, more elements of sparse variables are utilized at an iteration, thus the larger the value of $\alpha_{model}$.
	\tool has higher throughput than TF-PS for all the sparsity conditions.
	The fixed cost for dense variable communication is becoming more significant as the amount of data transfer for sparse variables reduces due to the small $\alpha_{model}$.
	Therefore, the biggest speedup of \tool compared to TF-PS is 3.42 when $\alpha_{model}$ is minimum.
\\

\section{Related Work}
\label{sec:rltwk}
\paragraph{Data Parallel Training on Existing DL Frameworks}
Existing DL frameworks, such as TensorFlow~\cite{tensorflow}, MXNet~\cite{mxnet} and PyTorch~\cite{pytorch}, support data parallel training with multiple machines and GPUs.
However, to the best of our knowledge, none of the existing frameworks consider the sparsity as an important factor of data parallel training, only supporting either the PS architecture or the AR architecture at one time.
Moreover, unlike \tool, most of the existing frameworks make users manually modify single-GPU code to be trainable in a distributed environment.

\sloppypar{
	For example, TensorFlow data parallelization APIs such as \texttt{SyncReplicasOptimizer}, \texttt{replica\_device\_setter}, \texttt{MonitoredTrainingSession} and \texttt{Server} are designed only for the PS architecture.
	Moreover, these APIs require additional modifications when converting a single-GPU graph to a distributed one, and users are still responsible for debugging if distributed training does not work correctly and efficiently.
}
To handle this issue, TensorFlow introduces a high-level \texttt{DistributionStrategy} API as an experimental feature, % TensorFlow v1.11 on September 2018.
which removes the manual modification process from users by converting a single-GPU code to a distributed version automatically.
However, even with such a high-level API, users must select which strategy to use among various strategies including \texttt{MirroredStrategy}, \texttt{CollectiveAllReduceStrategy} and \texttt{ParameterServerStrategy}, without any clue about the relationship between the model sparsity and training throughput.
Additionally, the programming model with \texttt{DistributionStrategy} is less flexible than the low-level data parallelization API to achieve automated distribution.
The current implementation of \texttt{DistributionStrategy}\footnote{TensorFlow v1.12, November 2018.} does not support synchronous multi-machine training with the PS architecture, input data sharding API for multi-machine training, and advanced performance optimizations that \tool provides.

MXNet~\cite{mxnet} supports data parallel training using a distributed key-value store for data synchronization between machines and GPUs, supporting only the PS architecture without considering the model sparsity.
In addition, a single-GPU code should be manually modified to pull variables and to push gradients using the store.
Moreover, it is impossible to improve communication efficiency by offloading some computations from a worker to servers with the key-value store.
PyTorch~\cite{pytorch} supports distributed training only with the AR architecture.
PyTorch provides APIs for constructing communication groups, averaging gradients, and adding aggregation methods for data parallel training.
Horovod~\cite{horovod} also provides an abstraction of efficient AR algorithms and implementations.
\tool also uses Horovod's MPI operators for TensorFlow including \texttt{HorovodAllreduceOp}.

\paragraph{Combining PS Architecture with Other Communication Mechanisms}
There exist other frameworks that try to improve performance by combining the PS architecture with other communication mechanisms.
MXNET-MPI~\cite{mxnetmpi} divides GPUs into multiple groups, where GPUs in the same group communicate using \texttt{AllReduce}/\texttt{Reduce} operations.
Each group then communicates with each other using the PS architecture.
For this new architecture, the paper introduces a new MPI Elastic SGD algorithm, which allows synchronous SGD methods within an MPI group and asynchronous SGD methods between groups to mitigate both the network contention problem in synchronous training and the staleness problem in asynchronous training.
The mixture of the PS architecture and \texttt{AllReduce}/\texttt{Reduce} operations is mainly used for controlling asynchrony for the new algorithm.
On the other hand, \tool combines the PS and AR architectures while maintaining the widely-used algorithm, synchronous SGD.
Moreover, since MXNET-MPI still uses collective communication within a group, it requires a larger amount of network transfer for handling sparse variables compared to \tool.

Poseidon~\cite{poseidon} combines the PS architecture and sufficient factor broadcasting (SFB) communication that uses peer-to-peer connections of workers.
SFB communicates sufficient factors of a gradient matrix for fully connected (FC) layers using its decomposability as two smaller vectors.
Even though Poseidon pursues a similar approach to choose an optimal training architecture based on the estimation of data transfer, it focuses on gradients of the FC layers, while \tool focuses on sparse variables and their gradients.

\paragraph{Model Parallel Training}
Model parallelism is another approach to deal with the large, sparse models.
In model parallelism, a single model is split across multiple GPUs, and each GPU computes only a part of the model.
A problem of model parallel training is underutilization of GPUs due to the small size of each fragment assigned to a GPU.
PipeDream~\cite{pipedream} addresses the problem using overlapped computation for multiple iterations.
However, the staleness caused by computing multiple iterations in parallel is getting significant if the number of GPUs increases.
Recently, hybrid strategies of model-parallelism and data parallelism~\cite{tofu, soybean} are introduced to find optimal parallelization methods by considering both sides, but they still need an efficient data parallel training to improve overall performance.

\paragraph{Increasing Variable Sparsity through Network Sparsification}
A dense model can be converted into a sparse model by employing pruning techniques~\cite{runtimepruning, autopruner} that are used to reduce the amount of computation, communication, and memory usage for both training and inference.
These techniques utilize different subsets of model variables for different inputs, making the variables sparse.
Quantization techniques~\cite{deepcompression, incquantization, quantization} change gradient tensors of dense variables into sparse formats by increasing the number of zero elements in the gradients.
Even when the model is intrinsically dense, by applying network pruning or quantization, we believe that \tool's hybrid architecture can outperform other frameworks that only utilize the PS or AR architecture.
We consider exploring this direction as future work.

\section{Conclusion} \label{sec:conc}
We present \tool, a framework that provides sparsity-aware data parallel training.
\tool introduces a hybrid approach that combines different training architectures according to the sparsity of variables to reduce the amount of network transfer.
\tool also proposes a method for partitioning sparse variables to maximize parallelism while maintaining low computation and communication overhead.
Its automatic graph transformation allows users to use their single-GPU program for training on a distributed environment while maintaining scalable performance.
We show that \tool achieves higher performance and scalability for sparse models compared to TensorFlow and Horovod in a cluster of 48 GPUs.
We open sourced \tool in the hope of facilitating users to take advantage of sparsity-aware data parallel training. 
\tool is publicly available at https://github.com/snuspl/parallax.	
\begin{acks}
We thank our shepherd Madan Musuvathi and the anonymous reviewers for their insightful comments.
This work was supported by the Institute for Information \& communications Technology Promotion(IITP) grant funded by the Korea government(MSIT) (No.2015-0-00221, Development of a Unified High-Performance Stack for Diverse Big Data Analytics), the ICT R\&D program of MSIT/IITP (No.2017-0-01772, Development of QA systems for Video Story Understanding to pass the Video Turing Test), and Samsung Advanced Institute of Technology.
\end{acks}

\balance
\bibliographystyle{ACM-Reference-Format}
\bibliography{parallax}

\end{document}